\documentclass[a4paper, amsfonts, amssymb, amsmath, reprint, showkeys, nofootinbib, twoside, , superscriptaddress]{revtex4-1}
\usepackage[english]{babel}
\usepackage[utf8]{inputenc}
\usepackage{amsmath,amssymb}
\usepackage[colorinlistoftodos, color=green!40, prependcaption]{todonotes}
\usepackage[pdftex, pdftitle={Article}, pdfauthor={Author}]{hyperref} 
\usepackage{color}
\definecolor{cblue}{RGB}{100,5,255}
\definecolor{cred}{RGB}{255,50,40} 
\definecolor{cgreen}{RGB}{40,255,40} 
     
\begin{document}
\title{Primordial black holes and gravitational waves in nonstandard cosmologies}

%
%

\author{Sukannya Bhattacharya}
    \email[]{sukannya@prl.res.in}
    \affiliation{Theoretical Physics Division, Physical Research Laboratory, Navrangpura, Ahmedabad - 380009, India.}
\author{Subhendra Mohanty}
    \email[]{mohanty@prl.res.in}
    \affiliation{Theoretical Physics Division, Physical Research Laboratory, Navrangpura, Ahmedabad - 380009, India.}
\author{Priyank Parashari}
    \email[]{parashari@prl.res.in}
    \affiliation{Theoretical Physics Division, Physical Research Laboratory, Navrangpura, Ahmedabad - 380009, India.}
    \affiliation{Indian Institute of Technology, Gandhinagar, 382355, India}


\begin{abstract}
For primordial black holes (PBHs) to form a considerable fraction of cold dark matter, the required amplitude of primordial scalar perturbations is quite large ($P_{\zeta}(k) \sim 10^{-2}$) if PBH is formed in radiation epoch. In alternate cosmological histories, where additional epoch of arbitrary equation of state precede radiation epoch, the dynamics of PBH formation and relevant mass ranges can be different leading to lower requirement of primordial power at smaller scales of inflation. Moreover, this alternate history can modify the predictions for the gravitational wave (GW) spectrum, which can be probed by upcoming GW observations. In this paper we show that an early kination epoch can lead to percent level abundance of PBH for a lower amplitude of $P_{\zeta}(k)$ as compared to PBH formation in a standard radiation epoch. Moreover, we calculate the effect of early kination epoch on the GW spectrum for first and second orders in perturbation theory which show enhancement in the amplitude of the GW spectrum in a kination epoch with respect to that in a standard radiation epoch.
\end{abstract}


\maketitle

\section{Introduction}
\label{intro}
The phenomenology of the early universe has entered a phase more rich than ever considering the numerous theoretical models in literature as well as the huge influx of data from several present and upcoming observations. The inflationary epoch, predicting quasiexponential expansion for $50$ - $60$ e-folds before hot big bang expansion, is constrained by the cosmic microwave background (CMB) surveys (latest is Planck 2018~\cite{Akrami:2018odb,Aghanim:2018eyx}). CMB observations predict the power spectrum of the primordial scalar fluctuations to be slightly red-tilted with amplitude $10^{-9}$. However, CMB only probes $10-15$ e-folds around the pivot scale $k_{\rm pivot}^{\rm CMB}=0.002~{\rm Mpc^{-1}}$. Lack of direct probe for the remaining $\sim 40$ e-folds leaves room for exploration where some models of inflation may have nonstandard predictions. A possible deviation from the red-tilt of the scalar power spectrum can take place for these smaller scales of inflation, which can lead to a rise in the scalar power spectrum. When these scales reenter the horizon, the large density fluctuations may collapse gravitationally to form primordial black holes (PBH).

PBHs are nonrelativistic and effectively collisionless and therefore have been proposed to be a dark matter (DM) candidate in early literature~\cite{Hawking:1971ei,Carr:1974nx,Carr:1975qj}. PBH as a candidate for DM has regained attention after the first detection of gravitational waves (GW) by Advanced LIGO/VIRGO in 2017~\cite{Abbott:2016blz,Abbott:2016nmj,TheLIGOScientific:2016pea,Abbott:2016bqf,Abbott:2017vtc,Abbott:2017gyy,Abbott:2017oio}. These GW interferometers have since detected several binary black hole mergers with mass of the initial components in the range $8-40M_{\odot}$. Stellar black holes, which form at the endpoints of stellar evolution under gravitational collapse after a supernova explosion, rarely have mass $\geq 10M_{\odot}$ as constrained from the x-ray emission from their accretion disks~\cite{Montero-Camacho:2019jte}. Moreover, given the high merging rates inferred by LIGO/VIRGO and the observed low effective spin point toward the possibility that the detected massive BH are of primordial origin~\cite{Fernandez:2019kyb}.

Although solar mass black holes are interesting given the detections of GW interferometers till date, the range of theoretically possible PBH mass $M$ and their abundance $f_{\rm PBH}(M)$ is constrained from many cosmological and astrophysical observations~\cite{Carr:2009jm}. Due to Hawking radiation, PBHs evaporate on a timescale $t_{\rm ev}=\frac{5120\pi G^2M^3}{\hslash c^4}$, so that PBHs of mass lower than $\sim 10^{15}~{\rm gm}$ have completely evaporated by now and produced $e^{\pm}$ pairs. Thus, PBHs with $M \lesssim 10^{15}~{\rm gm}$ are constrained by the observation of the Galactic Center $511~{\rm keV}$ gamma-ray line in SPI/INTEGRAL observatories~\cite{Churazov:2010wy,Siegert:2016ijv,Laha:2019ssq}. PBHs of mass $M\sim 10^{16}~{\rm gm}$ are on the verge of complete evaporation and thus have remnants from their Hawking evaporation in the extragalactic photon background, which is constrained with experiments such as Fermi Large Area Telescope~\cite{Carr:2009jm}. Hawking evaporation also leads to injection of positrons and neutrinos in the extragalactic background~\cite{Dasgupta:2019cae}, which are probed by SPI/INTEGRAL observatories and using diffuse supernova neutrino background at Super-Kamiokande~\cite{Bays:2011si} KamLAND~\cite{Collaboration:2011jza} and Borexino~\cite{Agostini:2019yuq}. 
Recent analysis of femtolensing surveys allow PBHs of mass $M\sim 10^{16}-10^{19}~{\rm gm}$ to contribute to a considerable fraction of the DM density~\cite{Barnacka:2012bm,Katz:2018zrn} whereas lack of microlensing events in EROS survey~\cite{Tisserand:2006zx} and MACHO collaboration~\cite{Allsman:2000kg} constrain the abundance of PBHs in the mass range $M\sim 10^{26}-10^{33}~{\rm gm}$~\cite{Niikura:2017zjd}. For massive PBHs, x-ray emission near PBH due to accretion of gas may modify recombination history and therefore affect spectral distortions and temperature anisotropies in CMB which constrain the abundance of PBHs with solar mass and above~\cite{Inoue:2017csr,Gaggero:2016dpq}. However, merging of subsolar mass PBHs in the late universe can lead to solar and super-solar mass PBHs that are observed now.
 Considering all these bounds, there is a very small range of masses $M\sim 10^{18}-10^{21}~{\rm gm}$ where PBH can still constitute the totality of DM.

Abundant PBH formation during radiation domination requires a large amplitude of the primordial curvature perturbation $P_{\zeta}(k)\simeq 0.02$, much larger than $P_{\zeta}(k)=2.1\times 10^{-9}$ constrained by CMB observations~\cite{Akrami:2018odb} at CMB relevant scales. Numerous studies have been done inspecting possible rise in power at smaller scales $k>k_{\rm CMB}$ for several theoretically and phenomenologically motivated models of inflation~\cite{Motohashi:2017kbs,Clesse:2015wea,Germani:2017bcs,Linde:2012bt,Bugaev:2013fya,Erfani:2015rqv,Cheng:2016qzb,Garcia-Bellido:2016dkw,Garcia-Bellido:2017aan,Garcia-Bellido:2017mdw,Ezquiaga:2017fvi,Bezrukov:2017dyv,Gao:2018pvq,Carr:2018nkm,Arya:2019wck}. For single field slow roll models of inflation, it is hard to achieve an ultra slow-roll ($\epsilon \sim 10^{-7}$) for smaller scales necessary for large power, which is sometimes achieved by including an inflection point~\cite{Garcia-Bellido:2017mdw} or a tiny bump~\cite{Mishra:2019pzq} in the inflaton potential. For multifield models of inflation and warm inflation, it is possible to achieve large power in the last $40$ e-folds of inflation involving the dynamics of fields other than inflaton. However, if PBHs are formed in an epoch prior to big bang nucleosynthesis (BBN) where the effective equation of state is different from $1/3$, then the critical density contrast and the background evolution are different which leads to a modification in the resulting PBH abundance. Ref.~\cite{Harada:2017fjm,Matsubara:2019qzv,Alabidi:2013wtp,Harada:2016mhb,Carr:2017edp,Hwang:2012bi} analyse formation of PBH in early matter dominated epochs whereas~\cite{Carr:2018nkm} explore PBH formation during reheating. However, in alternate cosmological histories where a nonstandard epoch precedes radiation epoch before BBN, the effective equation of state can even be $1/3<w\leq 1$. In this paper, we consider formation of PBH in such a nonstandard pre-BBN stiff epoch and analyse the resulting PBH abundance. 

During inflation, the quantum fluctuations of inflaton and metric perturbations result in fluctuations at all scales. Scalar, vector and tensor fluctuations evolve independent of each other in the first order of perturbation theory, however their evolutions are coupled when one considers second and higher order perturbations. The primordial tensor power spectrum at the first order is almost scale-independent with very small amplitude (constrained by CMB via tensor-to-scalar ratio $r$). The energy density $\Omega_{\rm GW}$ of the resulting stochastic gravitational wave (GW) background with inflation followed by a standard postinflationary evolution until now is beyond the reach of the sensitivities of current GW observations~\cite{Baumann:2007zm}. However, modification of pre-BBN history can modify the growth of the first order GW spectrum. The large amplitude of the curvature power spectra at small scales of inflation may give rise to large amplitude of tensor power spectrum at the second order of perturbation theory\footnote{for modifications in the GW spectra from primordial non-Gaussianities, check~\cite{Cai:2018dig}.}. The resulting large $\Omega_{\rm GW}$ at high frequencies $f=ck/2\pi$ can be within the sensitivity range proposed by the present and upcoming GW surveys such as pulsar time arrays (PTA), LISA and DECIGO~\cite{Lentati:2015qwp,Arzoumanian:2015liz,Lasky:2015lej,Bartolo:2016ami}. Ref.~\cite{Clesse:2018ogk} has explored second order GW spectra for several forms of primordial curvature power spectra with large amplitudes at small scales, which lead to abundant PBH formation in radiation epoch. Here we analyse the effect of a nonstandard pre-BBN epoch on first and second order GW spectra given a large amplitude of primordial curvature power spectrum at small scales (high frequencies).

In this paper, we particularly show the predictions for PBH and GW in an alternate cosmological history where a kination (kinetic energy domination) epoch ($w=1$) takes place between the end of inflation and the onset of radiation.\footnote{Here, we do not refer to the kinetic energy domination by slow roll violation at the end of inflation, since this is a very short duration and any PBH produced in this epoch will be of microscopic mass, therefore irrelevant to us.} The relevant quantities are derived in terms of a general $w$ in the range $1/3<w\leq 1$ for PBH analysis, but the final results are shown only for $w=1$ for clarity and simplicity. 

The rest of this paper is organized as follows. In Sec.~\ref{postreh01} we discuss the evolution of the energy density in the presence of a nonstandard pre-BBN epoch. In Sec.~\ref{pbhform} we derive the mechanism of PBH formation in such a nonstandard epoch and compare the resulting abundance to PBH formation in a purely radiation epoch. In Sec.~\ref{analysisPBH}, we consider different primordial curvature power spectra to arrive at an exact result for the PBH mass spectrum and abundance formed in an early kination epoch. We compare the results to those in radiation epoch resulting from the same set of power spectra. In Sec.~\ref{modGW}, we derive the GW spectra for the modified pre-BBN evolution. In Subsection~\ref{modGW1}, the enhancement of the first order GW spectrum is shown as a function of the scale of inflation $H_{\rm inf}$ and in Subsection~\ref{modGW2}, the second order GW spectrum is derived for an early kination epoch. In Sec.~\ref{discussion}, we discuss out results, conclude and comment on the possible future directions.


\section{Preradiation epoch of stiff domination}
\label{postreh01}
The observed abundance of light elements predicts that BBN must have taken place during radiation domination at $T_{\rm BBN}\simeq 1 {\rm MeV}$ and the evolution before that $T>T_{\rm BBN}$ remains inaccessible to direct observational techniques. Assuming instant reheating of the universe after inflation, the postinflationary pre-BBN epoch is considered to be radiation dominated in the standard big bang evolution. The simplest deviation from this scenario involves a reheating process where the inflaton field oscillates around the minimum of the potential for a few e-folds with an average equation of state $0<w<1/3$ (for $V(\phi) \propto \phi ^p$ with $p\leq 4$)~\cite{Lozanov:2016hid,Lozanov:2017hjm}\footnote{For implications of inflaton oscillation on GW spectra, check~\cite{Lozanov:2019ylm}. For implications of phase transitions and domain walls on PBH abundance and clustering, check~\cite{Khlopov:2008qy,Belotsky:2018wph,Khlopov:2004sc}.}. Many models of inflation predict additional epochs after reheating when an additional scalar field in the theory, which was inactive during inflation, dominates with an effectively matterlike equation of state ($w=0$) and decays, such as moduli domination at the end of several string theory inspired models of inflation~\cite{Conlon:2005jm,Maharana:2017fui,Bhattacharya:2017ysa}. The implications of a postinflationary scalar field domination on GW spectrum has been studied in~\cite{DEramo:2019tit}. However, in general there can be a postinflationary epoch with a stiff equation of state $1/3<w\leq 1$ which dominates the energy density before the onset of radiation domination~\cite{DiMarco:2018bnw}. Such a stiff dominated (SD) epoch may arise when a sterile field enters the postinflationary phase with dominant energy density that falls faster than radiation energy density with time. A particularly well-studied example is quintessence models of inflation~\cite{Peebles:1998qn,Ahmad:2019jbm} where the postinflationary epoch is dominated by the inflaton's kinetic energy with equation of state $w\approx1$ before transition into the radiation epoch sometime before BBN. As the universe expands, a  smooth transition from such an early kination epoch ($\rho \sim a^{-6}$) to radiation domination ($\rho \sim a^{-4}$) takes place at temperature $T_1$. 

We consider a stiff dominated epoch with a constant equation of state $w$ which dominates the energy density of the universe from the end of inflation until the onset of radiation domination at temperature $T_1$. We assume instantaneous transition from SD to radiation domination at $T_1$. The energy density at any time during this SD epoch is given in terms of the temperature at that time:
\begin{equation}
\rho_{\rm SD}(T)=\rho (T)=\rho_{\rm SD}(T_1)\bigg(\frac{a(T_1)}{a(T)}\bigg)^{3(1+w)}.\label{genrhoi}
\end{equation}
The energy density of SD and radiation can be equated at the transition such that:
\begin{equation}
\rho_{\rm SD}(T_1)=\rho_{\rm rad}(T_1)=\frac{\pi^2}{30}g_*(T_1)T_1^4,\label{rho1T1}
\end{equation}
where $g_*$ signifies the number of relativistic degrees of freedom. Moreover, conservation of entropy provides:
\begin{equation}
\frac{a(T_1)}{a(T)}=\bigg(\frac{g_s(T)}{g_s(T_1)}\bigg)^{1/3}\frac{T}{T_1},\label{scons01}
\end{equation}
where $g_s$ is the number of degrees of freedom contributing to the entropy of the universe. Therefore, 
\begin{equation}
\rho_{\rm SD}(T)=\frac{\pi^2}{30}g_*(T_1)\bigg(\frac{g_s(T)}{g_s(T_1)}\bigg)^{1+w}\bigg(\frac{T}{T_1}\bigg)^{3(1+w)}T_1^4.\label{rhophi1T}
\end{equation}
Evidently, this epoch has the maximum energy density at $T=T_{\rm reh}$ (or $T=T_{\rm infl.end}$ for an instantaneous reheating) and minimum possible energy density at $T_1=T_{\rm BBN}\simeq 1 {\rm MeV}$ equal to the radiation energy density at $T_1$.

If PBH formation takes place in such a SD epoch with $1/3<w\leq 1$, then the modified background evolution can affect the abundance of PBH corresponding to the modes entering the horizon in this epoch~\footnote{Check~\cite{Cotner:2017tir,Cotner:2016cvr} for PBH formation from Q-balls in the early universe.}. The evolution of the source-free first order tensor power spectrum is different in a SD epoch as compared to radiation dominated epoch~\cite{Figueroa:2019paj,Bernal:2019lpc} and can have an effective $w$-dependent blue-tilt. Moreover, the large amplitude of the primordial curvature power spectrum can source a large tensor power spectrum when second and higher order perturbation theory is considered. The resulting tensor power spectrum will evolve differently than in a pure radiation epoch after inflation and in such an alternate cosmological history leading to different $\Omega_{\rm GW, 0}h^2$ that can be probed by future GW interferometers.

In the rest of this paper, we present the calculations and basic notions for the dynamics of PBH formation and GW analysis in the presence of a postinflationary SD epoch. The PBH calculations and derivation for first order GW spectra are shown for a general stiff epoch with equation of state (e.o.s.) $w$, however, for the sake of clarity, while showing final results and plots, we resort particularly to an early kination epoch, $w=1$ (except for Fig.~\ref{gainplotmono} and Fig.~\ref{beta_PSPT}). For the derivations of second order GW spectra, we resort to fixing $w=1$ for the calculations as well as plots for the sake of simplicity.

\section{Formation of PBH}
\label{pbhform}
PBH can be formed in the early universe when the density fluctuations of high amplitude reenter the Hubble horizon at postinflationary epochs and collapse gravitationally below the Schwarzschild radius. The mass of the PBH at formation is:
\begin{equation}
M=\gamma M_H=\gamma \frac{4}{3}\pi(H^{-1})^3 \rho, \label{M01}
\end{equation}
where $M_H$ is the horizon mass, $\gamma $ describes the efficiency of the gravitational collapse~\cite{Carr:1975qj} and $H$ is the Hubble parameter at formation. PBH formation in the radiation epoch is studied in great detail in literature\footnote{See references in the 4th paragraph in Sec.~\ref{intro} and the references therein.}. However, as discussed in Sec.~\ref{intro}, the critical density contrast in radiation domination is such that the amplitude of primordial scalar power necessary for considerable PBH abundance during the radiation epoch is $\sim 10^{-2}$. In this section, we calculate the energy budget of PBHs formed during a stiff epoch with arbitrary $1/3<w\leq 1$ and analyze if considerable PBH abundance can be achieved with a lower amplitude of primordial scalar power.

The Friedmann equation for a stiff dominated epoch can be written as : $H^2=8\pi\rho_{\rm SD}(T)/3$ in $M_{\rm Pl}=1$ units. For a PBH formed in this epoch, the mass at formaion is:
\begin{eqnarray}
M(T)&=&\frac{\gamma}{2GH}=\frac{\gamma}{2G}\sqrt{\frac{3}{8\pi G}}\frac{1}{\sqrt{\rho_{\rm SD}(T)}}\nonumber \\
&=&\bigg(\frac{\gamma}{2G}\bigg)\bigg(\frac{\pi ^2 g_*(T_1)}{30}\times\frac{8\pi G}{3}\bigg)^{-\frac{1}{2}}\bigg(\frac{g_s(T_1)}{g_s(T)}\bigg)^{\frac{1+w}{2}}\nonumber \\ 
&&\times \bigg(\frac{T_1}{T}\bigg)^{\frac{3(1+w)}{2}}\frac{1}{T_1^2}. \label{M02}
\end{eqnarray}
The mass of a PBH formed in this epoch can be expressed in terms of the wavenumber $k$ of the fluctuations that entered the horizon\footnote{Throughout the manuscript, we consider that perturbations at mode $k=aH$ collapses immediately after entering horizon.} at temperature $T$. The Hubble parameter can be written as:
\begin{eqnarray}
\frac{H(T)}{H_{\rm eq}}&=&\frac{H(T)}{H(T_1)}\frac{H(T_1)}{H_{\rm eq}}\nonumber \\
&=&\bigg(\frac{a(T_1)}{a(T)}\bigg)^\frac{3(1+w)}{2}\bigg(\frac{a_{\rm eq}}{a(T_1)}\bigg)^2, \label{Ha_w}
\end{eqnarray}
where $H_{\rm eq}$ is the Hubble parameter at $T_{\rm eq}$ (matter-radiation equality).
Therefore, combining Eq.~\eqref{M01} and Eq.~\eqref{Ha_w}, the mass of PBH corresponding to mode $k$ is:
\begin{eqnarray}
M(k)&=&\bigg(\frac{\gamma}{2G}\bigg)\bigg(\frac{\pi ^2 g_*^{\rm eq}}{15}\times\frac{8\pi G}{3}\bigg)^{\frac{1}{3w+1}}\bigg(\frac{g_s^{\rm eq}}{g_s(T_1)}\bigg)^{\frac{3w-1}{3(3w+1)}}\nonumber \\
&& \times(a_{\rm eq}T_{\rm eq})^{\frac{3(1+w)}{3w+1}}T_1^{-\frac{3w-1}{3w+1}}k^{-\frac{3(1+w)}{3w+1}}. \label{Mk_w} 
\end{eqnarray}
Now, the present abundance of PBHs of mass $M$ over the logarithmic interval $d\ln M$ is:
\begin{equation}
f_{\rm PBH}(M) \equiv \frac{\Omega_{\rm PBH}(M)}{\Omega_c},\label{fM01}
\end{equation}
where $\Omega_{\rm PBH}(M)$ and $\Omega_c$ are the present energy densities (normalised, $\Omega = \rho/\rho_{\rm crit}$) for PBHs of mass in the range ($M$, $M+d\ln M$) and of cold dark matter (DM).
So, $f_{\rm PBH}(M)=1$ means that PBHs with a monochromatic mass distribution of mass $M$ constitute all of the cold dark matter. For a general mass distribution, $f_{\rm PBH}(M)$ can be determined neglecting the accretion and evaporation of the PBHs. Thus, for PBHs formed in a postreheating SD epoch of our consideration, the abundance will depend on the fractional contribution of PBHs at the end of that epoch, i.e., $\rho_{\rm PBH}(M)\vert_{T_1}$. The fraction of the universe collapsing into black holes of masses between $M$ and $M+d\ln M$ can be estimated in the Press-Schechter formalism of gravitational collapse as: 
\begin{eqnarray}
\beta(M)&\equiv &\frac{1}{\rho_{\rm tot}}\frac{d\rho_{\rm PBH (M)}}{d \ln M}=2\int_{\zeta_c}^{\infty}\frac{1}{\sqrt{2\pi}\sigma(M)}e^{-\frac{\zeta^2}{2\sigma(M)^2}}d\xi \nonumber \\
&=&{\rm erfc} \bigg(\frac{\zeta_c}{\sqrt{2}\sigma(M)}\bigg).\label{betadef}
\end{eqnarray}
Here, the critical curvature fluctuation $\zeta_c=\frac{(5+3w)}{2(1+w)}\delta _c$, where $\delta _c$ the critical density contrast of perturbations to gravitationally collapse and form PBH. 

$\sigma(M)$ is the variance of curvature fluctuations and is related to the primordial fluctuations through a window function $W(k,R)$. Generally $W(k,R)$ is considered to be a simple top hat function centered at $k_{\rm PBH}$ over which the power spectrum $P_{\zeta}(k)$ varies linearly~\cite{Young:2014ana}. The dependence of the critical density contrast on the equation of state has been studied analytically~\cite{Carr:1975qj} and numerically~\cite{Harada:2013epa} and we will consider the more precise numerical form~\cite{Harada:2013epa}\footnote{This expression for $\delta _c$ fails in a matter dominated epoch where it predicts the critical density contrast to be $\simeq 0$ such that all possible density perturbations entering the horizon can collapse~\cite{Kalaja:2019uju}. But, one can consider Eq.~\eqref{deltac_w} to be valid in a range $0 \ll w <1$ which is a superset of the SD equation of state considered here.}
\begin{equation}
\delta _c=\frac{3(1+w)}{(5+3w)}\sin ^2 \bigg(\frac{\pi \sqrt{w}}{(1+3w)}\bigg). \label{deltac_w}
\end{equation}
Considering $\sigma ^2(M)\simeq P_{\zeta}(k)$, the PBH mass fraction $\beta(M)$ can be written explicitly in terms of curvature perturbation spectrum $P_{\zeta}(k)$ as:
\begin{equation}
\beta(M)={\rm erfc} \bigg[\frac{3\times \sin ^2 \bigg(\frac{\pi \sqrt{w}}{(1+3w)}\bigg)}{2\sqrt{2P_{\zeta}(k)}}\bigg]. \label{beta_w}
\end{equation}
After formation, $\rho_{\rm PBH}$ grows as matter ($\sim a^{-3}$), whereas the background energy density grows as $\sim a^{-3(1+w)}$ until the temperature reaches $T_1$ and then as $\sim a^{-4}$ until matter-radiation equality at $T_{\rm eq}$. After equality, the PBH abundance becomes constant since both PBH and background evolve in similar manner ($\sim a^{-3}$).
Moreover, at the time of formation (PBHs of mass $M$ are formed at temperature $T$), only a fraction $\gamma \beta(M)$ of the total energy density turns into PBH. Therefore, considering all these factors,
\begin{eqnarray}
f_{\rm PBH}(M)&=&\frac{\Omega_{\rm PBH}(M)}{\Omega_c}=\frac{\rho_{\rm PBH}(M)}{\rho_c}\bigg\vert_{\rm eq}\nonumber \\
&=&\frac{\rho_{\rm PBH}(M)}{\rho_{\rm rad}}\bigg\vert_{\rm eq}\bigg(\frac{\Omega_m h^2}{\Omega_c h^2}\bigg)\nonumber \\
&=&\frac{\rho_{\rm PBH}(M)}{\rho_{\rm SD}}\bigg\vert_{T_1}\bigg(\frac{a(T_{\rm eq})}{a(T_1)}\bigg)\bigg(\frac{\Omega_m h^2}{\Omega_c h^2}\bigg)\nonumber \\
&=&\frac{\rho_{\rm PBH}(M)}{\rho_{\rm SD}}\bigg\vert_{T}\bigg(\frac{a(T_1)}{a(T)}\bigg)^{3w}\bigg(\frac{a(T_{\rm eq})}{a(T_1)}\bigg)\bigg(\frac{\Omega_m h^2}{\Omega_c h^2}\bigg).\nonumber \\
\label{fM02}
\end{eqnarray}
Here, in the second line we have used the condition that matter and radiation energy density are equal at $T_{\rm eq}$. Similarly, in the third line, we have used $\rho_{\rm SD}(T_1)=\rho_{\rm rad}(T_1)$. Using the conservation of entropy density at $T_{\rm eq}$, $T_1$ and $T$,
\begin{equation}
a(T_{\rm eq})^3T_{\rm eq}^3g_s(T_{\rm eq})=a(T_1)^3T_1^3g_s(T_1)=a(T)^3T^3g_s(T),\label{s-conserve}
\end{equation} 
Eq.~\eqref{fM02} can be written in terms of temperature $T$ of PBH formation as:
\begin{eqnarray}
f_{\rm PBH}(M)&=&\gamma \beta(M) \bigg(\frac{g_s(T)}{g_s(T_1)}\bigg)^w\bigg(\frac{g_s(T_1)}{g_s(T_{\rm eq})}\bigg)^{1/3}\bigg(\frac{T}{T_1}\bigg)^{3w}\nonumber \\
&&\times \bigg(\frac{T_1}{T_{\rm eq}}\bigg)\bigg(\frac{\Omega_m h^2}{\Omega_c h^2}\bigg). \label{fM03}
\end{eqnarray}
Substituting $T$ from Eq.~\eqref{M02} in Eq.~\eqref{fM03}, the abundance can be expressed in terms of $M$ as:
\begin{eqnarray}
f_{\rm PBH}(M)&=&\frac{\gamma}{T_{\rm eq}}\bigg(\frac{g_s(T_1)}{g_s(T_{\rm eq})}\bigg)^{\frac{1}{3}}\bigg(\frac{\Omega_m h^2}{\Omega_c h^2}\bigg)\bigg(\frac{3}{8\pi G}\frac{30}{\pi ^2g_*(T_1)}\bigg)^{\frac{w}{1+w}}\nonumber \\
&& \times \bigg(\frac{\gamma}{2G}\bigg)^{\frac{2w}{1+w}}T_1^{\frac{1-3w}{1+w}}\beta (M)M^{-\frac{2w}{1+w}}.\label{fM04}
\end{eqnarray}
Therefore, for an extended mass distribution, the total abundance of PBH is:
\begin{eqnarray}
f_{\rm PBH}^{\rm tot}&=&\int f_{\rm PBH}(M) d\ln M \nonumber \\
&=&\mathcal{C}_1 \mathcal{C}_2(w) T_1^{\frac{1-3w}{1+w}}\int \beta (M) M^{-\frac{2w}{1+w}} d\ln M , \label{fPBHtot}
\end{eqnarray}
where
\begin{eqnarray}
\mathcal{C}_1&=&\frac{\gamma}{T_{\rm eq}}\bigg(\frac{g_s(T_1)}{g_s(T_{\rm eq})}\bigg)^{1/3}\bigg(\frac{\Omega_m h^2}{\Omega_c h^2}\bigg)~{~ ~ \rm and}~\nonumber \\
\mathcal{C}_2(w)&=&\bigg(\frac{3}{8\pi G}\frac{30}{\pi ^2g_*(T_1)}\bigg(\frac{\gamma}{2G}\bigg)^2\bigg)^{\frac{w}{1+w}}.\label{C2wval}
\end{eqnarray}
Eq.~\eqref{fM03} can express the abundance of PBHs of mass $M$ produced in a radiation epoch with the limit $w\rightarrow1/3$ and $T_1\rightarrow T_{\rm eq}$, so that,
\begin{equation}
f_{\rm PBH}^{\rm rad}(M)=\gamma \beta ^{\rm rad}(M) \bigg(\frac{g_s(T)}{g_s(T_{\rm eq})}\bigg)^{1/3}\bigg(\frac{T}{T_{\rm eq}}\bigg)\bigg(\frac{\Omega_m h^2}{\Omega_c h^2}\bigg),
\label{fM03_rad}
\end{equation}
\begin{figure}[htbp]
\includegraphics[width=0.49\textwidth]{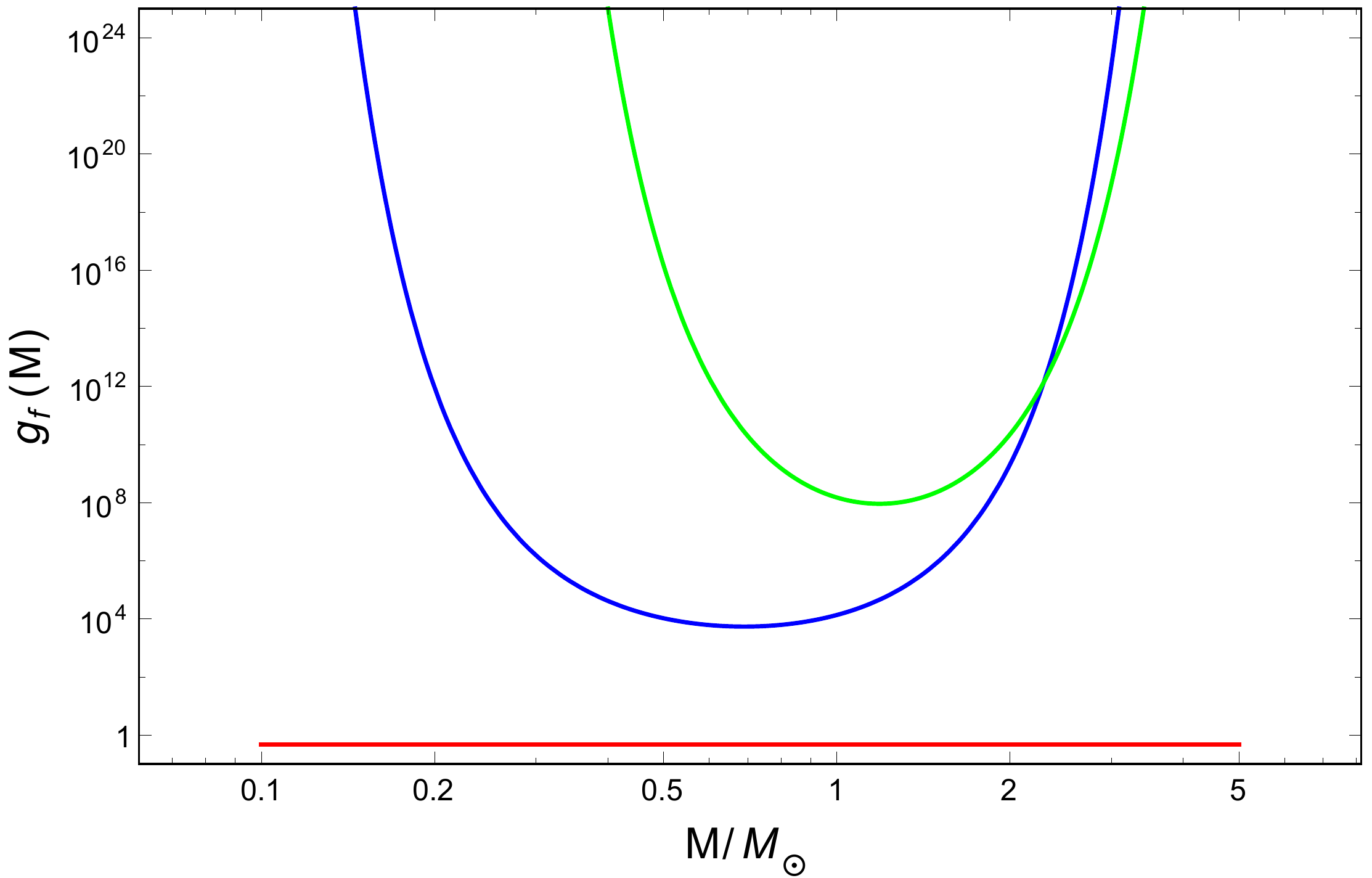}\caption{Gain $g_f(M)$ as a function of $M$ with $T_1=10$ MeV for a $P_{\zeta}(k)$ with a single sharp peak at $k_p=2\times 10^6$ Mpc$^{-1}$. \textit{Red}, \textit{blue} and \textit{green} curves are for $w=1/3$, $w=2/3$ and $w=1$ respectively.}\label{gainplotmono}
\end{figure}
where $\beta ^{\rm rad}(M)={\rm erfc} \bigg(\frac{1.02}{\sqrt{2P_{\zeta}(k)}}\bigg)$, since $\zeta _c=1.02$ in a radiation dominated epoch. Thus, for any given primordial power $P_{\zeta}(k)$, the gain in the abundance of PBH in a stiff dominated epoch over a radiation dominated epoch is:
\begin{eqnarray}
g_f(M)&=&\frac{f_{\rm PBH}(M)}{f_{\rm PBH}^{\rm rad}(M)}\nonumber \\
      &=&\frac{\beta (M)}{\beta ^{\rm rad}(M)}\frac{g_s(T)^{w-1/3}}{g_s(T_1)^{w-1}g_{s}(T_{\rm eq})^{2/3}}\bigg(\frac{T}{T_1}\bigg)^{3w-1}.\label{gain01}
\end{eqnarray}
%
For all $1/3<w\leq 1$, we find $g_f(M)>1$. Therefore, a lower amplitude of the curvature power spectrum than that required for PBH formation in radiation epoch can lead to a considerable percent level abundance of PBH formed in a SD epoch. For heavier PBHs (i.e. $T\simeq T_1$), the gain depends on the ratio of $\beta (M)$.

For instance, in Fig.~\ref{gainplotmono}, we plot $g_f(M)$ as a function of $M$ with $T_1=10$ MeV for different $w$ for a curvature power spectrum with a single sharp peak at a particular scale $k_p$ . The exact form of $P_{\zeta}(k)$ is same as in Eq.~\eqref{pow3} with $P_p=0.01$, $\sigma_p=0.5$ and $k_p=2\times 10^6$ Mpc$^{-1}$. This plot clearly shows that the gain in PBH abundance is large in presence of a pre-BBN SD epoch and maximum for $w=1$. For a particular value of $k_p$, corresponding PBH mass range and profile depend on $w$ via Eq.~\eqref{Mk_w}. Therefore, $w=2/3$ (blue) and $w=1$ (green) correspond to two different mass ranges, which is evident in Fig.~\ref{gainplotmono}.
\begin{figure}[htbp]
\includegraphics[width=0.49\textwidth]{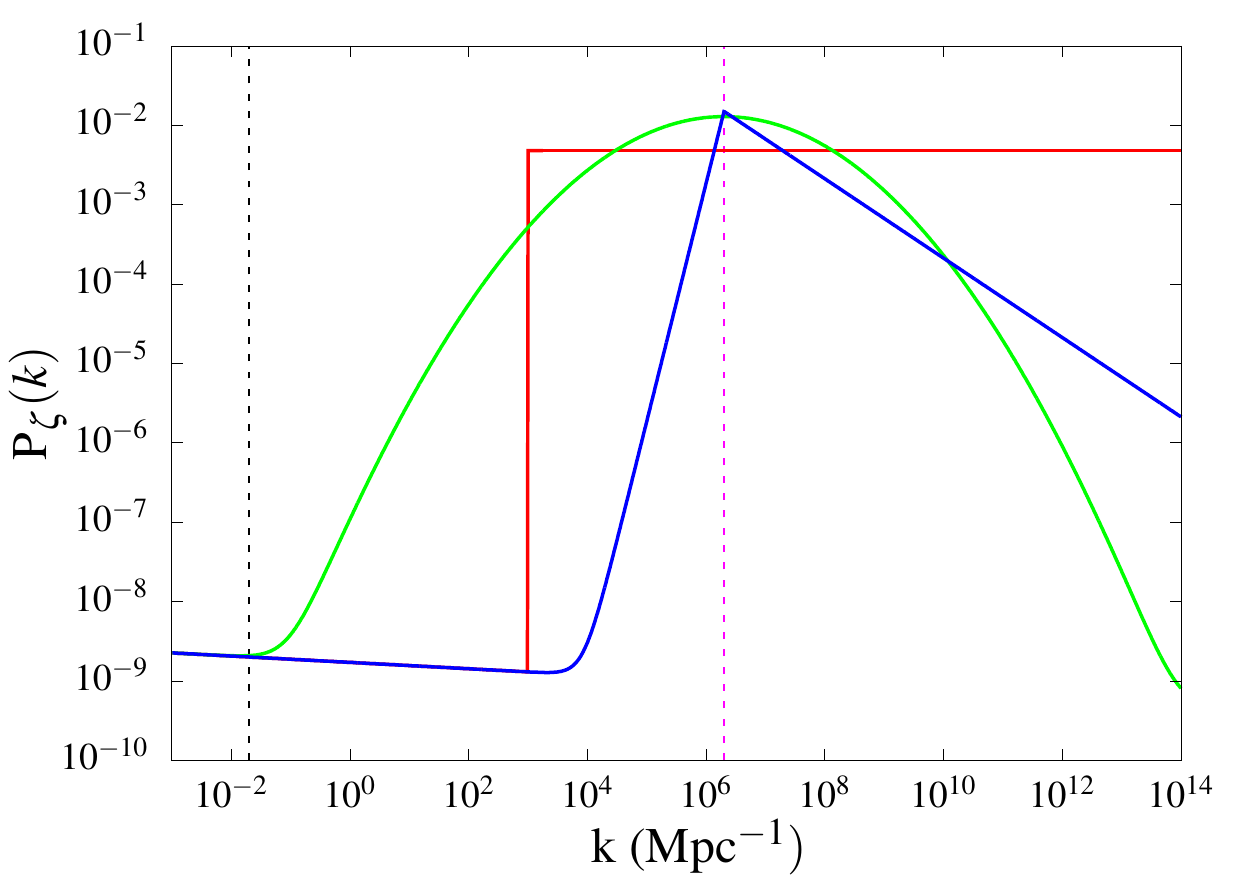}\caption{Curvature power spectrum for different cases: \textit{Red:} Scale invariant, \textit{Blue:} Broken power law, \textit{Green:} Gaussian power spectrum. Black dashed line points to the CMB pivot scale $k_{*}=0.002 ~{\rm Mpc^{-1}}$ and magenta dashed line signifies $k_p=2\times 10^6 ~{\rm Mpc^{-1}}$ for production of near-solar mass PBHs.}
\end{figure}
\begin{figure}[htbp]
\includegraphics[width=0.5\textwidth]{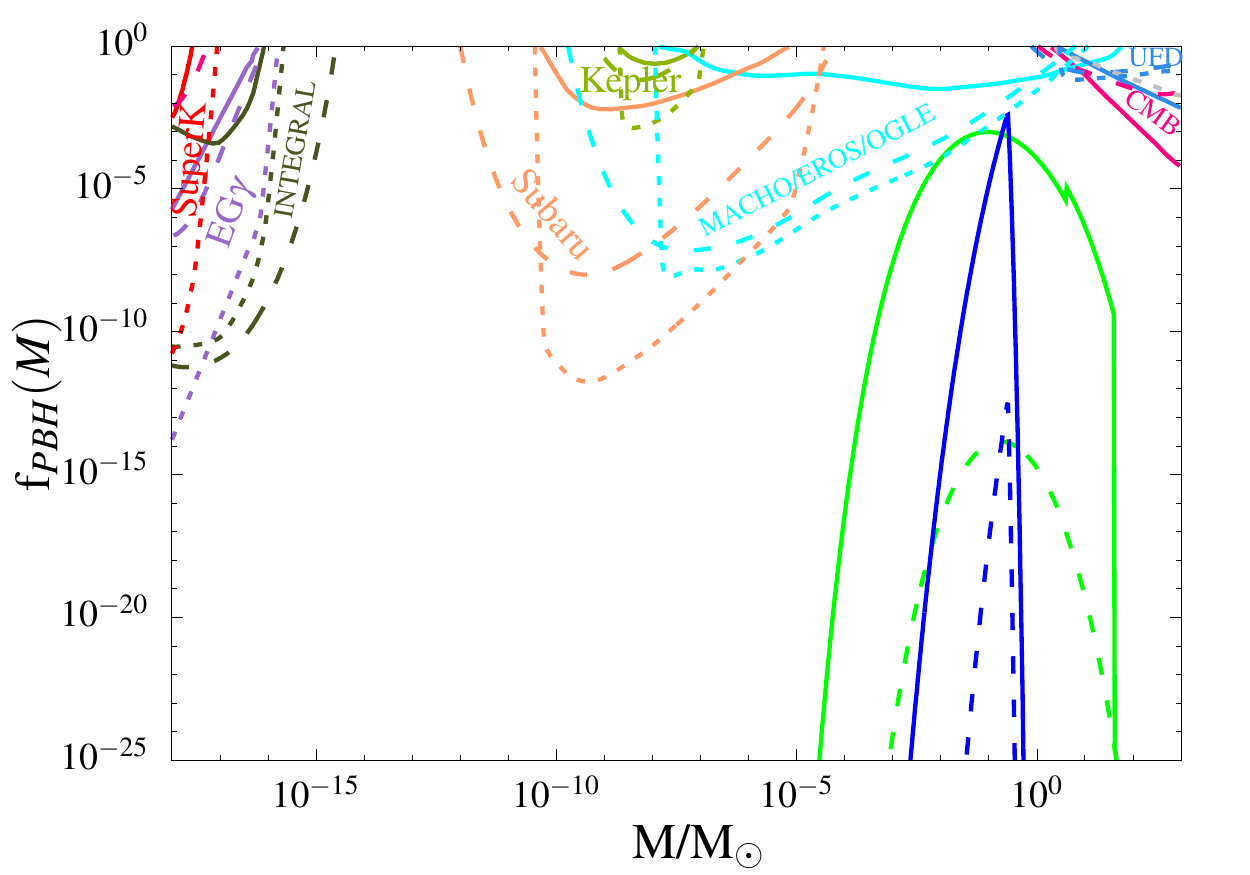}
\caption{Abundance of PBH for broken power law power spectrum (blue) and for gaussian power spectrum (green) for PBH formation in a pure radiation dominated epoch (dashed lines) and in a kination epoch before radiation domination (solid lines). For PBH formation in a kination epoch, we considered that PBH contributes to $\sim 10\%$ fraction in total DM energy density. Constraints on the PBH mass spectra are shown for different cosmological and astrophysical observations where solid curves represent constraints for a monochromatic mass spectrum, long-dashed curves for a Gaussian power spectrum and short-dashed curves for the broken power law power spectrum (details in text). The peak scale here is $k_p=2\times 10^{6}~{\rm Mpc^{-1}}$.}
\label{fig_solar}
\end{figure}
\section{Analysis for different primordial power spectra}
\label{analysisPBH}
In this section, we study three different types of primordial curvature power spectra from a phenomenological approach with similar motivations as~\cite{Clesse:2018ogk}. Apart from the first case, which provides a clear idea about requirement of smaller power in a SD epoch, other two cases are motivated by theories of inflation. In all the cases, $A_s$ and $n_s$ refer to CMB constrained primordial scalar amplitude and scalar spectral index respectively.
\subsection{Scale invariant power spectrum}
Here, we consider a constant amplitude of the primordial curvature power spectrum for all modes beyond $k_p$ such that:
\begin{equation}
P_{\zeta}(k)=A_s\bigg(\frac{k}{k_*}\bigg)^{n_s-1}+P_p\Theta(k-k_p),\label{pow1}
\end{equation}
where $P_p$ is the power at all scales with $k>k_p$. We emphasize that such a constant amplitude of the scalar power at small scales is not motivated by theoretical considerations and eventually leads to a considerable abundance of PBH for all the masses below $M(k_p)$. However, we calculate the abundance for this case for simplicity and completion, although we do not include them in the plots.
\subsection{Broken power law power spectrum}
In various scenarios of the early universe where PBH is produced from domain walls or vacuum bubbles~\cite{Deng:2016vzb,Deng:2017uwc}, the relevant primordial curvature power spectrum has a broken power law form such as:
\begin{equation}
P_{\zeta}(k)=\left\{ \begin{array}{l l} A_s\bigg(\frac{k}{k_*}\bigg)^{n_s-1}+P_p\bigg(\frac{k}{k_p}\bigg)^m & \ \quad k<k_p,\\
A_s\bigg(\frac{k}{k_*}\bigg)^{n_s-1}+P_p\bigg(\frac{k}{k_p}\bigg)^{-n} & \ \quad k\geq k_p \end{array}\right.
\label{pow2}
\end{equation}
We consider the case where $m=3$ and $n=0.5$.
\subsection{Gaussian power spectrum}
In many models of smooth waterfall hybrid inflation~\cite{Clesse:2015wea} and several inflection point models of inflation~\cite{Germani:2017bcs}, the potential features a plateau for a few e-folds before the end of inflation. This plateau regime of the potential can lead to a peak in the curvature power spectrum which, at the simplest approach, can be written as a Gaussian power spectrum of the following form:
\begin{equation}
P_{\zeta}(k)=A_s\bigg(\frac{k}{k_*}\bigg)^{n_s-1}+P_p \exp\bigg[-\frac{(N_k-N_p)^2}{2\sigma _p^2}\bigg], \label{pow3}
\end{equation}
where $N_k=\ln(a(k)/a_{\rm end})$ is the number of e-folds before the end of inflation when the mode $k$ exits the horizon such that $N_p=\ln(a(k_p)/a_{\rm end})$ and we consider $\sigma_p=3$.

\begin{figure}[htbp]
\includegraphics[width=0.5\textwidth]{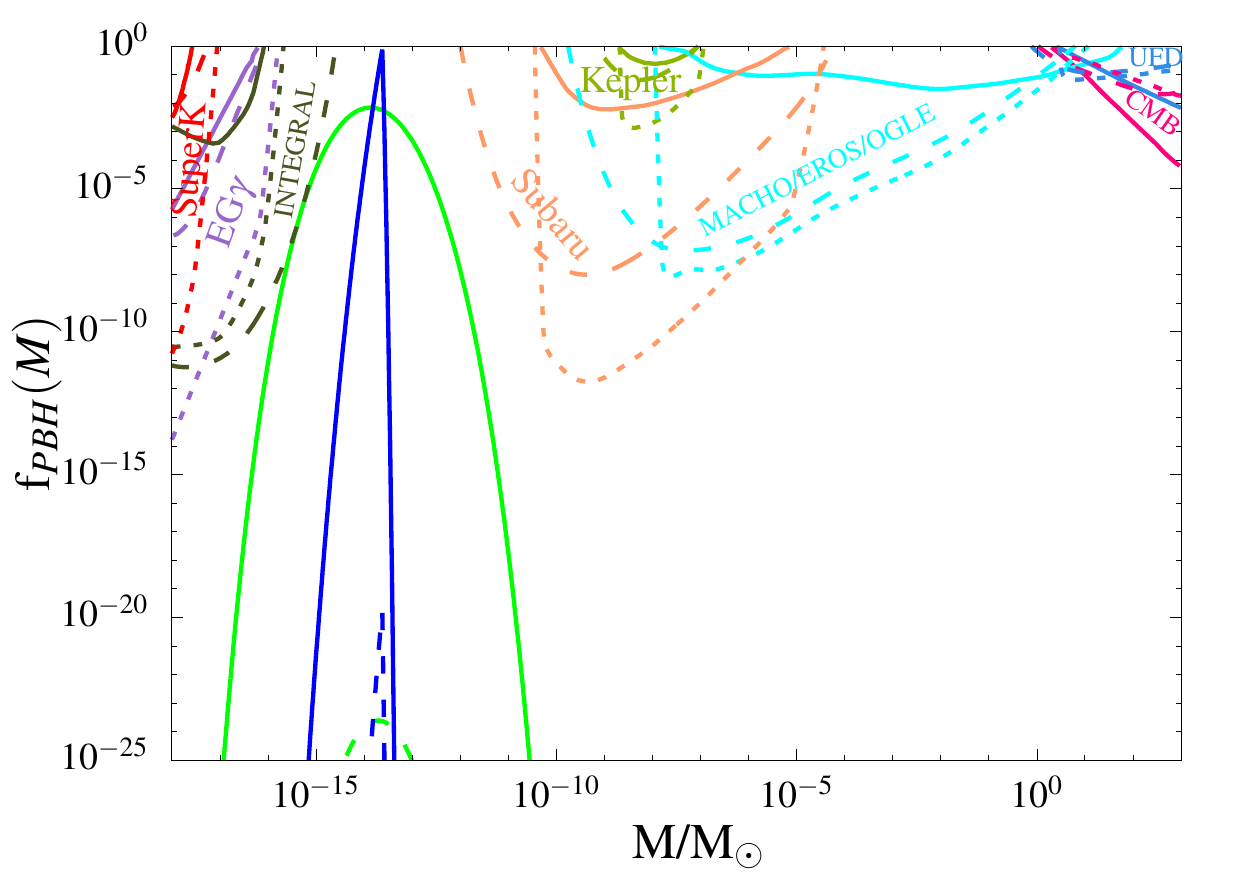}
\caption{Abundance of PBH for broken power law power spectrum (blue) and for gaussian power specturm (green) for PBH formation in a pure radiation dominated epoch (dashed lines) and in a kination epoch before radiation domination (solid lines). For PBH formation in a kination epoch, we considered that PBHs contributes to $\sim 10\%$ fraction in total DM energy density. Constraints on the PBH mass spectra are shown for different cosmological and astrophysical observations and the curves have same specifications as Fig.~\ref{fig_solar}. The peak scale here is $k_p=6\times 10^{12}~{\rm Mpc^{-1}}$.}
\label{fig_LISA}
\end{figure}
\subsection*{Analysis} 
The above three primordial power spectra are shown in Figure 2 for a specific value of the peak scale $k_p = 2 \times 10^6$ Mpc$^{-1}$. For each of the three power spectra, the first term represents a red-tilted power spectrum at the CMB relevant scales, as constrained by Planck~\cite{Akrami:2018odb}, with $A_s=2.1\times 10^{-9}$, $n_s=0.9649$ and $k_*=0.002~{\rm Mpc^{-1}}$. The second part of $P_{\zeta}(k)$, which reaches maximum amplitude at $k=k_p$ for the broken power law and Gaussian power spectrum, is relevant for leading to abundant PBH formation. The peak scale $k_p$ shifts the features in the power spectrum along the mass-axis of PBH whereas, theoretically, $k_p$ is related to the dynamics of the underlying inflation model\footnote{As an example, modifying $k_p$ for the Gaussian power spectrum for an waterfall-hybrid inflation model may translate to changing the starting point of the mild waterfall phase. Detailed discussion on the exact implication for inflation models in beyond the scope of this paper.}. 

The peak mode $k_p=2\times 10^6~{\rm Mpc^{-1}}$ is relevant for formation of near-solar mass PBH, which is explored in~\cite{Clesse:2018ogk} for PBH formation in a radiation epoch. However, in this paper we explore the implications of $k_p=2\times 10^6~{\rm Mpc^{-1}}$ for PBH formed in an early kination epoch. Fig.~\ref{fig_solar} shows the PBH mass spectrum for the broken power law (blue) and Gaussian (green) power spectra with respect to several astrophysical and cosmological constraints. This figure clearly points to a higher abundance $f_{\rm PBH}(M)$ for PBH formation in an early kination epoch (solid lines) than formation in a radiation epoch (dashed lines). 
Here, we have considered $T_1= 10~{\rm MeV}$ that corresponds to PBH of $M\sim 10M_{\odot}$. Therefore, with $k_p=2\times 10^6~{\rm Mpc^{-1}}$ the feature of the primordial power spectra spans through some part of the radiation epoch after SD as well. Thus, the mass spectrum in Fig.~\ref{fig_solar} makes a sharp transition from large abundance in a kination epoch to a much smaller abundance in a radiation epoch at $M=M(T_1)$. Evidently, such a kination epoch is not helpful to have larger abundance for PBHs with $M \gg M_{\odot}$, but such extremely massive PBHs can be formed at later epochs via merging of two or more solar or subsolar mass PBHs, which are abundant in this scenario.
\begin{table}[]
\caption{Comparison of required amplitude of curvature power spectra for a SD epoch with $w=1$ and for a radiation epoch to achieve a $\sim 10\%$ abundance of DM.}
\begin{center}

\resizebox{0.48\textwidth}{!}{
\begin{tabular}{|c|c|c|c|c|}
\hline 
$k_p ({~\rm Mpc^{-1}})$ & $w$ & Scale-inv $P_p$ & Broken Power Law $P_p$ & Gaussian $P_p$\\
\hline
\hline 
$2\times10^6 $ & $1/3$ & 0.021 & 0.0275 & 0.025\\
\hline 
$2\times10^6 $ & $1$ & 0.0048 & 0.0113 & 0.0105\\
\hline
\hline
$6\times 10^{12} $ & $1/3$ & 0.013 & 0.016 & 0.0163\\
\hline
$6 \times 10^{12} $ & $1$ & 0.0048 & 0.0067 & 0.006\\
\hline
\end{tabular}}
\label{Table1}
\end{center}
\end{table}

The same analysis has been carried out for $k_p=6\times 10^{12}~{\rm Mpc^{-1}}$
for which we determine the amplitude of the curvature power spectrum required for a considerable abundance of PBHs of relevant masses around $10^{-17}M_{\odot}<M<10^{-10}M_{\odot}$ for the Gaussian power spectrum and $10^{-16}M_{\odot}<M<10^{-13}M_{\odot}$ for the broken power law power spectrum. Interestingly, this mass range includes a very small mass window where it is still allowed by observations to achieve totality of DM from PBHs. Moreover, this $k_p$ corresponds to the frequency $f\simeq 0.01~{\rm Hz}$, which is in the maximum sensitivity regime for future LISA mission (details in Sec.\ref{modGW2}). In this case, the transition to radiation domination (RD) is taken to be at $T_1=10~{\rm MeV}$ as well. Fig.~\ref{fig_LISA} shows the mass spectrum for PBH for the two different power spectra and clearly points to a gain in abundance for a kination epoch compared to a standard RD epoch. 

In~\cite{Carr:2017jsz}, it is argued that the observational constraints on an extended PBH mass spectrum is different from (but related to) the constraints for a monochromatic mass function. For an extended mass spectrum $\psi (M)$ defined as: $f_{\rm PBH}^{\rm tot}\equiv \int \psi (M) dM$, any observable $A[\psi (M)]$ can be expanded as: 
\begin{eqnarray}
A[\psi (M)]&=&A_0+\int dM \psi (M) K_1(M) \nonumber \\
&+& \int dM_1 dM_2 \psi (M_1) \psi (M_2) K_2(M_1, M_2),\label{obsA}
\end{eqnarray}
where the functions $K_j$ depend on the nature of the observable quantity and underlying physics associated with it. If a certain measurement puts a bound on this observable such that $A[\psi(M)]\leq A_{\rm exp}$, then for a monochromatic mass spectrum peaking at $M=M_c$, the maximum observationally allowed abundance from Eq.~\eqref{obsA} is
\begin{equation}
f_{\rm max}(M_c)\equiv \frac{A_{\rm exp}-A_0}{K_1(M_c)}. \label{obs2}
\end{equation}
Therefore, for an extended mass distribution, this observation would put the constraint:
\begin{equation}
\int dM \frac{\psi (M)}{f_{\rm max}(M_c)} \leq 1. \label{obs3}
\end{equation}
Comparing the definition of $\psi (M)$ with our definition of the mass spectrum in Eq.~\eqref{fPBHtot}, $\psi(M)\equiv \frac{f_{\rm PBH}(M)}{M}$. Performing the above integration in Eq.~\eqref{obs3} in the mass range relevant to the particular experiment with observable $A[\psi (M)]$ would then lead to expressing the maximum allowed $f_{\rm PBH}$ from observation in terms of the parameters in the mass spectrum. This is done here for the Gaussian curvature power spectrum of Eq.~\eqref{pow3} where the mass spectrum $f_{\rm PBH}(M)$ depends on $P_p$, $k_p$ and $\sigma _p$ and for the broken power-law power spectrum  of Eq.~\eqref{pow2}, where $f_{\rm PBH}(M)$ depends on $P_p$, $k_p$, $m$ and $n$. In both the cases, $k_p$ corresponds to the mass where the mass spectra peak (check Eq.~\eqref{Mk_w} with $w=1$) and the values of other parameters of the spectra are kept at fixed values as earlier. Thus, varying $M(k)$ results in the constraints on $f_{\rm PBH}$ for specific observations. In Fig.~\ref{fig_solar}, and Fig.~\ref{fig_LISA}, these observational constraints are shown for monochromatic (solid lines), Gaussian power spectrum (long-dashed lines) and broken power-law power spectrum (short-dashed lines) for several observations: SuperKamiokande in red, extragalactic $\gamma$-rays in violet, INTEGRAL in dark green, Subaru in brown, Kepler in light green, microlensing (MACHO/EROS/OGLE) in cyan, ultrafaint dwarfs (UFD) in blue and from CMB constraints in pink~\cite{Ali-Haimoud:2016mbv,Allsman:2000kg,Tisserand:2006zx,Carr:2009jm,Griest:2013esa,Brandt:2016aco,Niikura:2017zjd,Smyth:2019whb,Boudaud:2018hqb,Dasgupta:2019cae}.

The lower requirement for the amplitude of primordial power $P_p$ is shown in the Table~\ref{Table1} which shows a significant improvement for $w=1$ epoch with respect to the radiation epoch, specifically for the case $k_p=6\times 10^{12}~{\rm Mpc^{-1}}$. We discuss the implications of these results in more details in Sec.~\ref{discussion}.
%
%
%
 
\section{Modification of the Gravitational Wave spectra}
\label{modGW}
Given a model of inflation, stochastic background of GW arise inevitably from the primordial tensor fluctuations. Standard single field slow roll inflation models lead to the tensor power spectrum $\Delta ^2_{h,{\rm inf}}(k)\equiv \frac{k^3}{2\pi^2}\vert h^{\rm inf}_k\vert^2=A_t\times (k/k_*)^{n_t}$ with $A_t\sim r\times A_s$ and a slight blue-tilt $n_t$. The tensor spectral index $n_t$ has a lower bound from the observed tensor to scalar ratio $r<0.056$ (at $95\%$ confidence limit (C.L.) from Planck~\cite{Akrami:2018odb}) and the single field consistency relation $r=-8n_t$. The evolution of the primordial tensor fluctuations is source-free in the first order of perturbation theory. The energy density of GW today is determined as
\begin{equation}
\Omega_{\rm GW,0}(k)=\frac{k^2\Delta ^2_h(\eta_0,k)}{12a_0^2H_0^2},\label{omtot}
\end{equation}
where $\Delta ^2_h(\eta_0,k)=T(k,\eta_0)\times \Delta ^2_{h,{\rm inf}}(k)$ is the tensor power spectrum today. 

The transfer function $T(k,\eta_0)$ depends on the evolution of the background from the time of horizon reentry of a mode $k$ until the time of observation $\eta_0$. For standard cosmology where the modes reenter during radiation domination, the transfer function gives a constant growth for all modes $k$ since the GW grows in the same way as the background ($\sim a^{0}$) in a radiation epoch. However, in alternate cosmological histories with an additional SD epoch, the transfer function behaves differently owing to the departure of $w$ from $w=1/3$ and the relative growth of GW is $\sim a^{-4+3(1+w)}$.
\subsection{First order perturbation theory}
\label{modGW1}
In the first order of perturbations, after horizon reentry, the evolution of the source free tensor fluctuations depend on the e.o.s.~$w$ and the temperature of transition from SD to radiation: $T_1$. The primordial tensor perturbations also depend on the scale of inflation $H_{\rm inf}$. Here, for simplicity, we have considered a scale-independent primordial tensor power spectrum, i.e., $n_t\simeq 0$. Tracking the growth of GW originating at the first order tensor fluctuations from horizon reentry until today, the GW energy density is~\cite{Figueroa:2019paj,Bernal:2019lpc}:
\begin{eqnarray}
\Omega_{\rm GW,0}^{(1)}(k)&=&\frac{\Omega_{\rm rad, 0}}{12\pi^2}\bigg(\frac{g_{*,k}}{g_{s,k}}\bigg)\bigg(\frac{g_{s,0}}{g_{s,k}}\bigg)^{4/3}\bigg(\frac{H_{\rm inf}}{M_{\rm Pl}}\bigg)^2\nonumber \\
&& \times \frac{\Gamma ^2 (\alpha +1/2)}{2^{2(1-\alpha}\alpha ^{2\alpha}\Gamma ^2(3/2)}\mathcal{W}(\kappa)\kappa^{2(1-\alpha)},\label{GW_1_gen}
\end{eqnarray}
where $\alpha=\frac{2}{1+3w}$, $\kappa=\frac{k}{k(T_1)}=\frac{f}{f(T_1)}$ and
\begin{equation}
\mathcal{W}(\kappa)= \frac{\pi \alpha}{2\kappa}\bigg[\bigg(\kappa J_{\alpha+1/2}(\kappa)-J_{\alpha-1/2}(\kappa)\bigg)^2+\kappa^2J^2 _{\alpha-1/2}(\kappa)\bigg],\label{Bessel_GW1}
\end{equation}
where $J_{i}$ is the Bessel function of order $i$.
\begin{figure}[htbp]
\centering{
\includegraphics[width=0.48\textwidth]{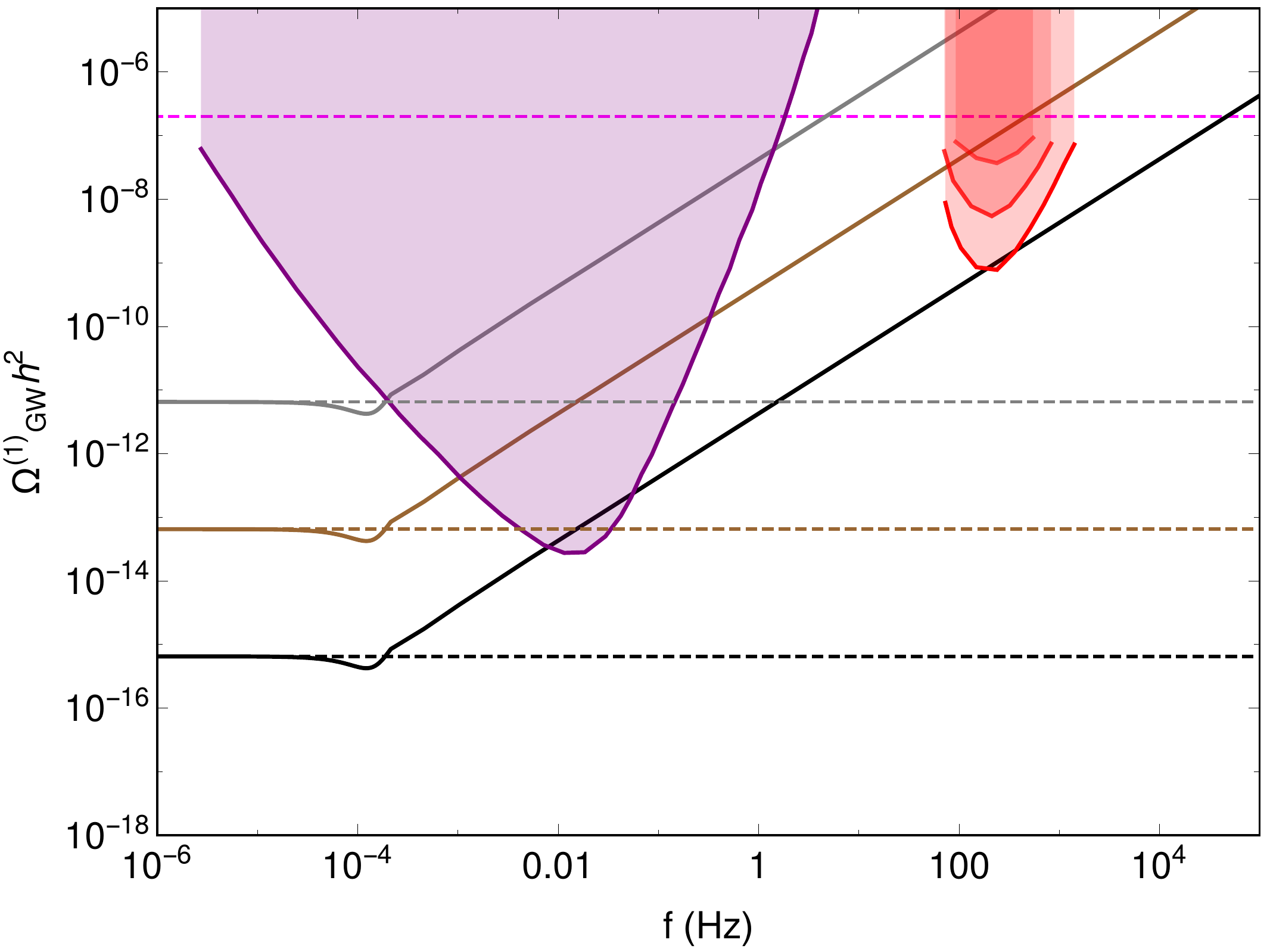}
}
\caption{GW spectrum arising in the first order of perturbation theory for an early kination epoch $w=1$ with $T_1=490~{\rm GeV}$ for $H_{\rm inf}=10^{12}$, $10^{13}$ and $10^{14}~{\rm GeV}$ (black, brown and grey curves respectively). Corresponding evolutions in a pure radiation epoch are shown by dashed black, brown and grey curves for respective cases. Bounds from CMB, LIGO and LISA are shown in magenta dashed line, red shaded region and purple shaded region respectively.}\label{GW_order1}
\end{figure}
The resulting GW spectra for an early kination epoch $w=1$ for $ T_1=490~{\rm GeV}$ is shown in Fig.~\ref{GW_order1} for different $H_{\rm inf}$ values. The bounds on GW from LIGO O1, O2 and O5 (future) runs (red curves) and future LISA (purple curve) constrain the GW spectra. CMB puts a bound on the GW energy density fraction since GW contribute to the radiation energy density. If GWs are overproduced, they will affect the expansion rate of the universe during CMB decoupling, which may not be allowed from the constraint on the extra radiation component obtained from CMB. For the homogeneous initial condition of GW, CMB puts an upper bound~\cite{Caprini:2018mtu,Smith:2006nk,Sendra:2012wh,Pagano:2015hma} on fraction of GW energy density as: $ \Omega_{\rm GW} h^2 \leq 2\times10^{-7}$ which is shown by the magenta dashed line. Particularly, LIGO O2 bound (middle red curve) signifies that if there is an early kination epoch after inflation until $T_1\simeq 490~{\rm GeV}$ then the scale of inflation is below $H_{\rm inf}\simeq 5\times 10^{12}~{\rm GeV}$.

\subsection{Second order perturbation theory}
\label{modGW2}
We have closely followed~\cite{Baumann:2007zm} to calculate the second order GW spectrum induced from the first order scalar perturbations. Evolution of second order tensor perturbations $h_k$ is given as
\begin{equation}
\label{equ:hhh}
h_k'' + 2 {\cal H} h_k' + k^2 h_k=  {\cal S}(\mathbf{k}, \eta)\, , 
\end{equation}
where the source term ${\cal S}(\mathbf{k}, \eta)$ is
\begin{equation}
 \label{equ:source}
 {\cal S}(\mathbf{k}, \eta) = \int {\rm d}^3 \mathbf{p} \,  
p^2 [1-\mu^2] f(\mathbf{k}, \mathbf{p}, 
\eta) \, \psi_{\mathbf{k-p}} \psi_{\mathbf{p}}\, .
\end{equation}
\begin{figure}[] 
   \centering
   \includegraphics[width=0.48\textwidth]{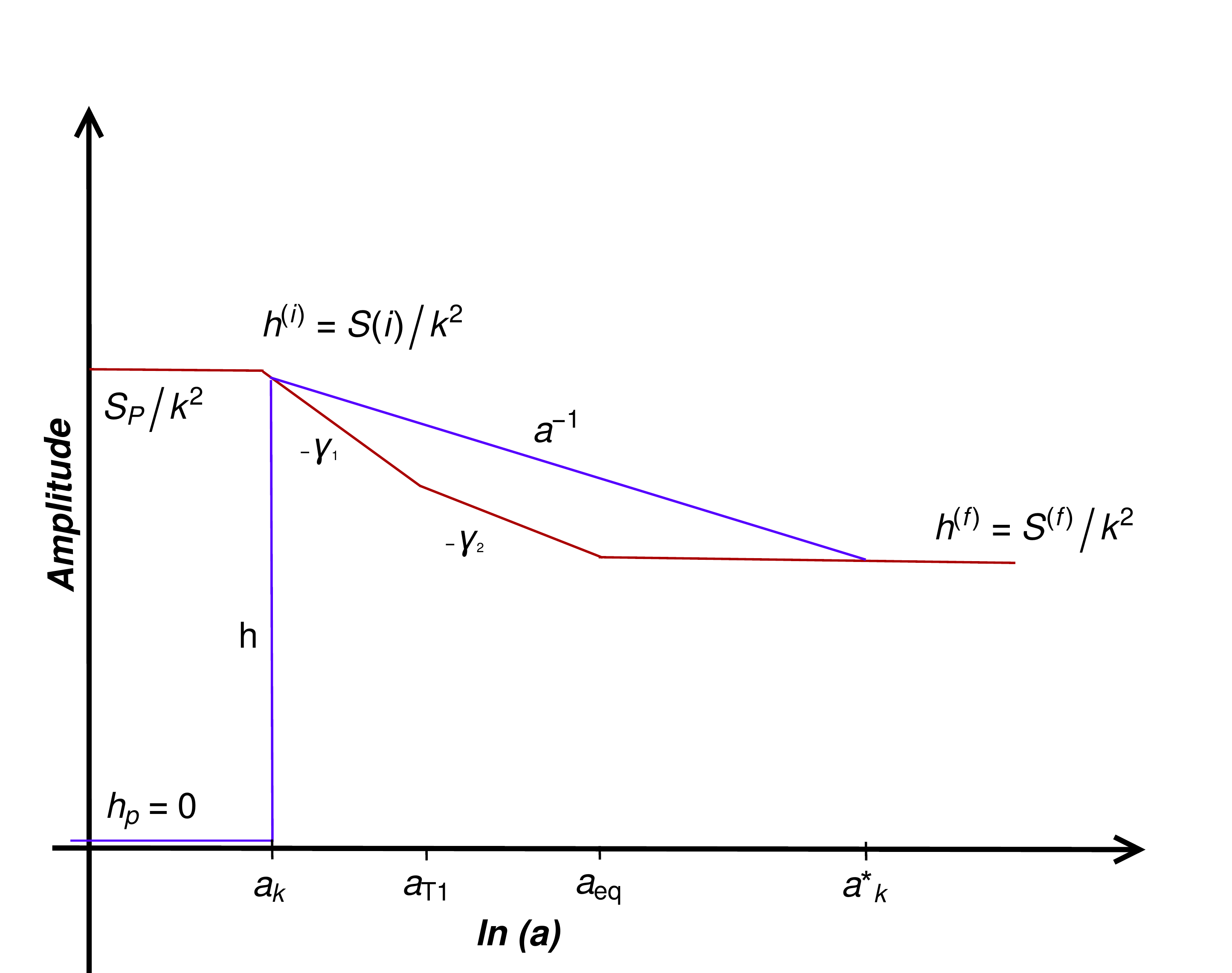} 
   \caption{Evolution of scalar source and induced gravitational waves throughout different epochs after horizon reentry. The source evolves as $a^{-\gamma _1}$ and $a^{-\gamma _2}$ in the kination and radiation epoch respectively and becomes constant at the beginning of matter domination. The induced tensor fluctuations evolve as $a^{-1}$ during kination and radiation era and settle down at some later epoch (matter domination) at the constant value set by the source term. The modes relevant to our analysis correspond to very high frequencies, and therefore they have not yet settled down completely at the constant value of the source term. }
   \label{fig:evol}
\end{figure}
In this expression, $\psi_{\mathbf{p}}$ is the primordial scalar perturbation and $\mu = \frac{\mathbf{k} 
\cdot \mathbf{p}}{k p}$ with $f(\mathbf{k}, \mathbf{p}, \eta)$ given as:
\begin{eqnarray}
f(\mathbf{k}, \mathbf{p}, \eta) =\frac{8}{3(1+w)} \big[ (5+3 w) \Phi(|\mathbf{k-p}| \eta) \Phi(| \mathbf{p}| \eta) \nonumber \\ + 2 \left( 2 \eta \Phi(|\mathbf{k-p}|\eta) +  \eta^2 \Phi'(|\mathbf{k-p}|\eta) \right)
  \Phi'(|\mathbf{p}|\eta)\big].\label{equ:f}
\end{eqnarray}
Here $\Phi(| \mathbf{ p}| \eta)$ is the first order scalar transfer function, which is related to the Bardeen potential $\Phi_{\mathbf{p}}(\eta)$ and primordial fluctuations $\psi_{\mathbf{p}}$ as:
\begin{equation}
\Phi_{\mathbf{p}}(\eta) = \Phi(p\eta) \, \psi_{\mathbf{p}}.
\end{equation}
Now, in order to calculate the evolution of the source term we need to calculate the first order scalar transfer function $\Phi(p\eta)$. In Appendix~\ref{sec:sca_trans}, we have derived the expression for $\Phi(p\eta)$ to be
\begin{equation}
\label{equ:Phiw}
\Phi(p \eta) = \left\{ \begin{array}{cc}\frac{2}{p \eta} J_1(p \eta)\,  ~~~~{\rm for}~~  \eta < 
\eta_1\\  
\frac{3\sqrt{2}}{p \eta \pi} \Bigl[ A(p) j_{1}\left(\frac{p \eta}{\sqrt{3}}\right) + B(p) 
y_{1}\left(\frac{p \eta}{\sqrt{3}}\right)\Bigr]\, ~ {\rm for}~~\eta_1 \leq \eta < \eta_{\rm{eq}} 
\end{array} \right . 
\end{equation}
where $\eta_1 \equiv \eta _{T_1}$, $J_1(p \eta)$ is the Bessel function and $j_{1}\left(\frac{p \eta}{\sqrt{3}}\right)$ and $y_{1}\left(\frac{p \eta}{\sqrt{3}}\right)$ are the spherical Bessel function of first and second type. Expressions for $A(p)$ and $B(p)$ are given in Appendix~\ref{sec:sca_trans}.

We are working under the approximation that GWs are generated instantaneously as the primordial curvature perturbations of a particular mode reenter the horizon. After horizon reentry, the evolution of that mode is governed by the transfer function. Therefore, the second order tensor transfer function $t(k,\eta)$ is defined as 
\begin{equation}
h_k(\eta) = t(k,\eta) h_k^{\rm (i)}\, ,
\end{equation}
where $h_k^{\rm (i)}$ is the initial amplitude of the primordial GW. At the horizon entry, we can neglect the time derivative terms in Eq.~\eqref{equ:hhh} to calculate $h_k^{\rm (i)}$, therefore
\begin{equation}
 \label{equ:hi}
h_k^{\rm (i)} \sim \frac{1}{k^2} {\cal S}^{\rm (i)}\, ,
\end{equation}
The tensor power spectrum for any mode at the time of horizon crossing is defined as
\begin{equation}
 \label{equ:pi}
P_h^{\rm (i)}(k,\eta_{\rm i}(k)) = \frac{k^3}{2 \pi^2} \langle (h_k^{\rm 
(i)})^2\rangle \, ,
\end{equation}
where $\eta_{\rm i}(k)$ is the time when mode $k$ enters the horizon. After horizon entry, the mode evolves according to the transfer function. Thus, the second order power spectrum at any time can be written as:
\begin{equation}
P_h(k,\eta) =t(k,\eta)^2 P_h^{\rm (i)}(k,\eta_{\rm i}(k)) .
\end{equation}
 Now the expression for power spectrum at horizon crossing $P_h^{\rm (i)}(k,\eta_{\rm i}(k))$ can be derived using the Eq.s~\eqref{equ:source},Eq.~\eqref{equ:hi} and Eq.~\eqref{equ:pi}. After some straightforward algebra, we arrive at 
\begin{widetext}
\begin{equation}
P_h^{\rm (i)}(k,\eta_{\rm i}(k)) = \frac{k^3}{2 \pi^2} \langle (h_k^{({\rm i})})^2 \rangle 
\sim
\frac{1}{2 \pi^2 k} \int {\rm d}^3 \mathbf{p} \, p^4 (1 - \mu^2)^2 \Phi^2(p \eta_{\rm i}) \Phi^2(|\mathbf{k}-
\mathbf{p}| \eta_{\rm i})
\frac{P(p)}{p^3} \frac{P(|\mathbf{k}-\mathbf{p}|)}{|\mathbf{k}-\mathbf{p}|^3}\,  \label{equ:phi0}
\end{equation}
\end{widetext}
where $P(p)$ is the primordial scalar power spectrum. In this work we have used two different scalar power spectra, as given in Eq.s~\eqref{pow2} and~\eqref{pow3} in section~\ref{analysisPBH}, to the calculate second order tensor power spectra. In the next sub-section, we will derive the expression for the second order tensor transfer function.

\subsubsection{Tensor transfer function}
Second order tensor transfer function depends on the evolution of the source term ${\cal S}$. In Fig.~\ref{fig:evol} we have shown how the evolution of the source term and the amplitude of GW changes if we have an extra period of stiff domination. Now let us consider a mode $k=a_k H$, which enters the horizon during SD period. At horizon entry, the GW amplitude immediately becomes equal to the source term as given in Eq.~\eqref{equ:hi}. If the source term decays as $a^{\gamma_1(k)}$ during the SD era and as $a^{\gamma_2(k)}$ during the RD era that follows, then 
\begin{equation}
\label{equ:S_rat}
\frac{{\cal S}^{\rm (f)}}{{\cal S}^{\rm (i)}} = \left( \frac{a_k}{a_{\rm 
T1}}\right)^{\gamma_1(k)}\left( \frac{a_{\rm T1}}{a_{\rm 
eq}}\right)^{\gamma_2(k)}\, ,
\end{equation}
where $a_{\rm T1}$ is the scale factor at the time when SD ends. It was shown in Ref.~\cite{Baumann:2007zm} that source term, ${\cal S} \approx \frac{1}{\eta^2} \frac{1}{(k \eta)^2} \propto \frac{1}{a^4}$  for the modes entering during RD epoch. We have also used the similar arguments to calculate the asymptotic values of both $\gamma_1(k)$ and $\gamma_2(k)$ for the modes which enter the horizon during SD era and found that the asymptotic value of $\gamma_1(k)$ is $6$, whereas  $\gamma_2(k)$ has asymptotic value equal to 2 for these modes (see Appendix~\ref{sec:tens_trans} for detailed derivation). Moreover, the upper limit on $\gamma_2(k)$ for the modes which enter the horizon during RD is found to be equal to $4$, which is in agreement with~\cite{Baumann:2007zm}. Hence, the source term decay faster in SD era than in RD era.
On the other hand GW amplitude $h$ just redshifts as $a^{-1}$ till the time it becomes equal to source term at time $a^*_{k}$. Therefore, we have
\begin{equation}
\frac{h^{\rm (f)}}{h^{\rm (i)}} = \frac{a_k}{a^*_k} \approx \frac{{\cal S}^{\rm 
(f)}}{{\cal S}^{\rm (i)}} =  \left( \frac{a_k}{a_{\rm 
T1}}\right)^{\gamma_1(k)}\left( \frac{a_{\rm T1}}{a_{\rm 
eq}}\right)^{\gamma_2(k)}\, \, ,
\end{equation}
Now let us define a sufficiently small scale $k_c$ for a fixed time $\eta$, such that modes with $k > k_c(\eta)$ have never settled down by the time $\eta$, they just keep on redshifting as $a^{-1}$. Thus, we find the critical scale $k_c(\eta)$ for the modes which enter during SD era
\begin{equation}
\label{eq:kc}
\hspace{-0.75cm}
k_c(\eta) =k_{T1}\left[\bigg(\frac{a(\eta)}{a_{T1}}\bigg)^{\frac{\gamma_1(k)}{1-\gamma_1(k)}}\bigg(\frac{a_{T1}}{a_{eq}}\bigg)^{{\frac{\gamma_2(k)}{1-\gamma_2(k)}}}
\bigg(\frac{k_{eq}}{k_{T1}}\bigg)^{-\gamma_2(k)}\right]^{\frac{-2}{\gamma_2(k)}} ,
\end{equation}
and  for the modes which enter during RD era
\begin{equation}
\label{eq:kcrd}
k_c(\eta) =\left( \frac{a(\eta)}{a_{\rm eq}}\right)^{\frac{1}{(\gamma_1(k)-1)}} k_{\rm eq}\,,
\end{equation}
where $k_{\rm T1}$ and $k_{\rm eq}$ are the scales corresponding to $a_{\rm T1}$ and $a_{\rm eq}$ respectively.
Now for the modes with $k> k_c(\eta)$ and $k>k_{\rm T1}$, transfer function is given as:
\begin{equation}\label{tran:sd}
t(k, \eta) = \frac{a_k}{a(\eta)} = \left(\frac{k_{\rm T1}}{k}\right)^{1/2} \frac{k_{\rm eq}}{k_{\rm T1}} \,\frac{a_{\rm eq}}{a(\eta)}, \hspace{0.5cm} k > k_c(\eta)\, ,~~ k>k_{\rm T1} .
\end{equation}
However, for the modes with $k_{\rm T1} < k < k_c(\eta) $, transfer function is
\begin{equation}
t(k, \eta) =  \left(\frac{k_{\rm T1}}{k}\right)^{\gamma_1(k)/2} \left(\frac{k_{\rm eq}}{k_{\rm T1}}\right)^{\gamma_2(k)}\ , \hspace{1cm} k_{\rm T1} \leq k < k_c(\eta) \,.
\end{equation}
\begin{figure}
\centering{
\includegraphics[width=0.5\textwidth]{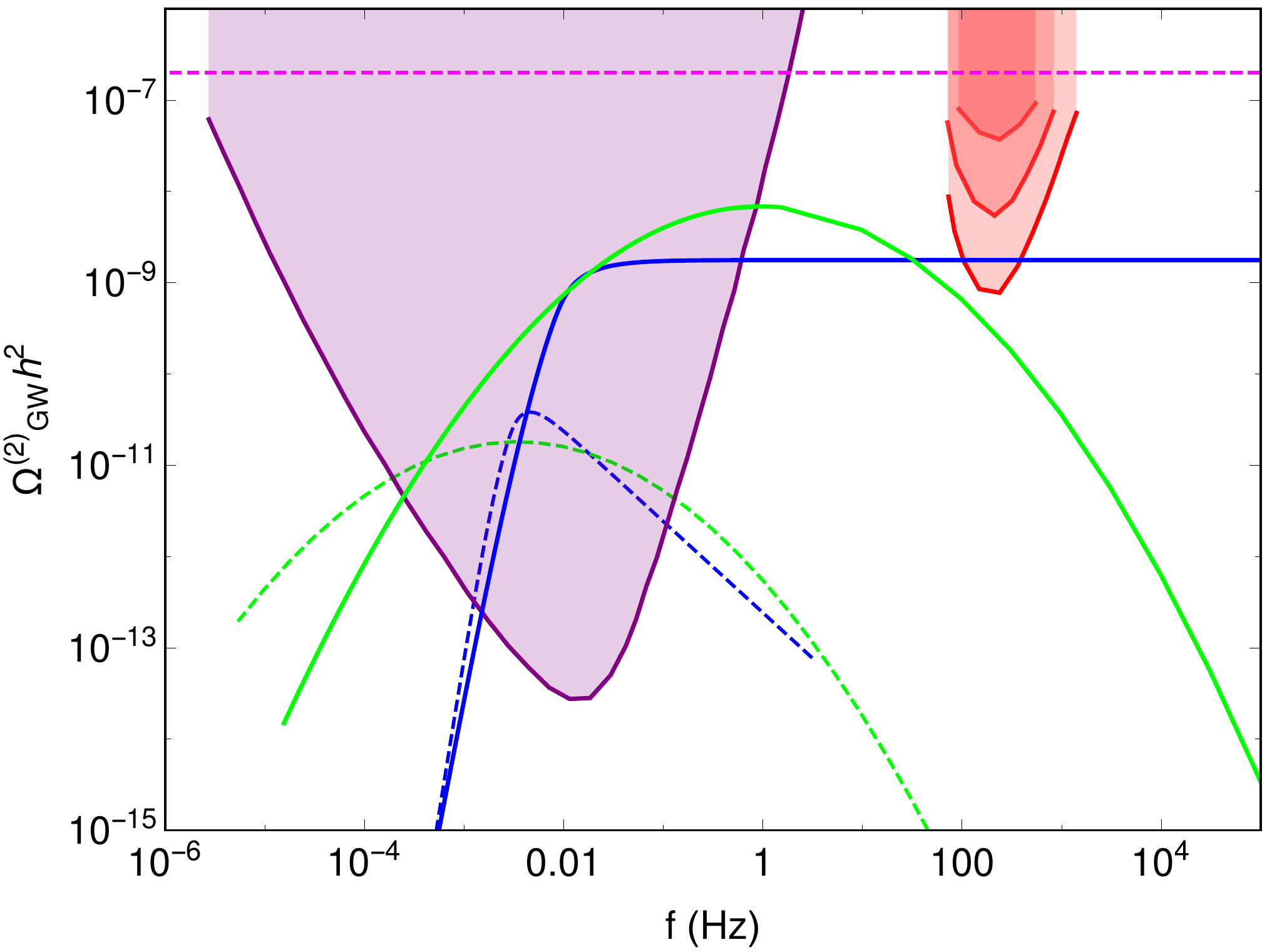}
\caption{Second order GW spectrum for the Gaussian primordial power spectrum (green curve) and broken power law power spectrum (blue curve) for the cases when relevant modes enter the horizon in a kination epoch (solid lines) and in a radiation epoch (dashed line). The spectrum in the kination epoch is enhanced for both the cases over that in the radiation epoch. Bounds from CMB, LIGO and LISA are shown in magenta dashed line, red shaded region and purple shaded region respectively.}\label{fig:2ndGW}
}
\end{figure}
\begin{figure}
\centering{
\includegraphics[width=0.5\textwidth]{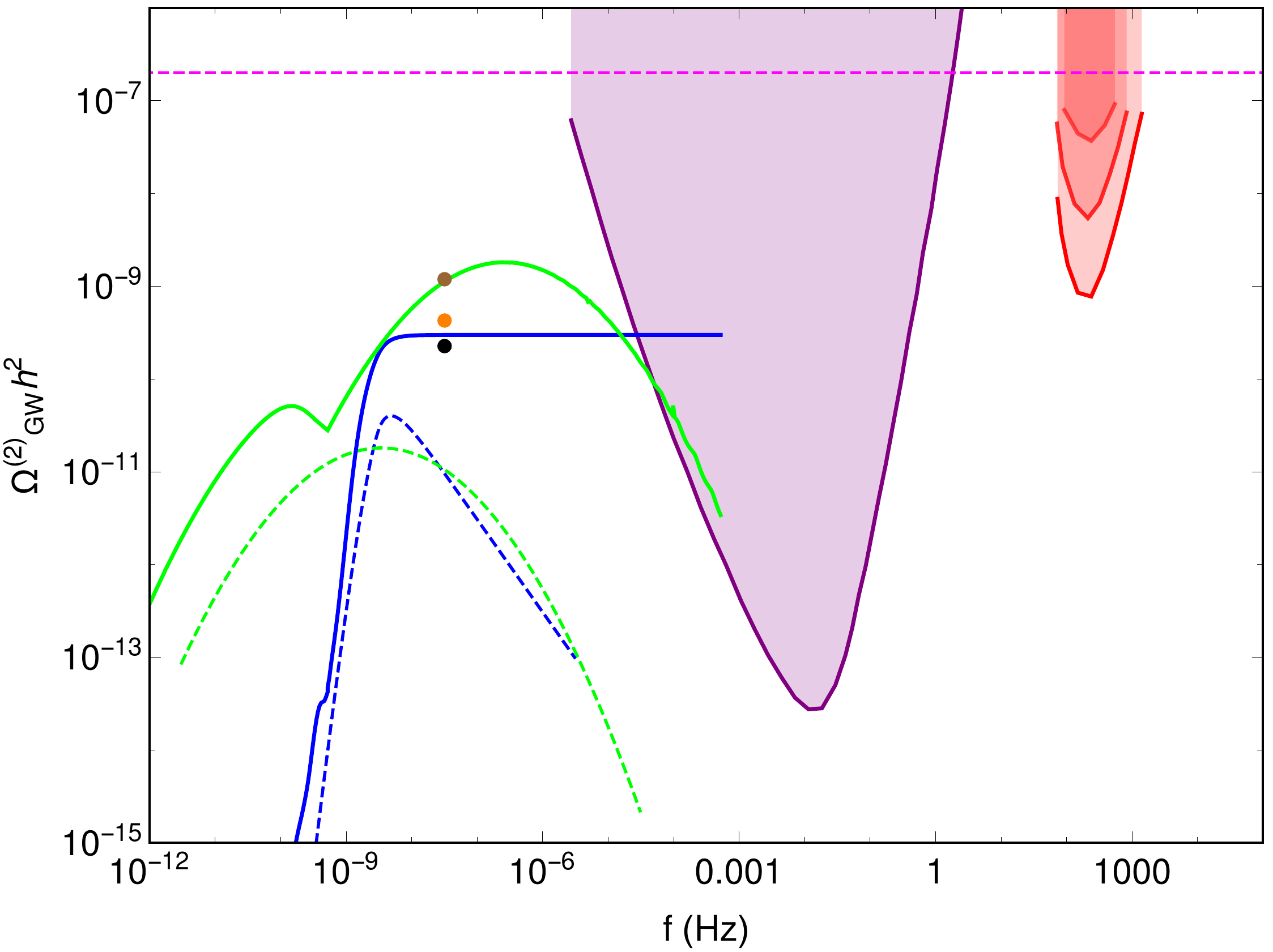}
\caption{Second order GW spectrum for the Gaussian primordial power spectrum (green curve) and broken power law power spectrum (blue curve) is shown here. The spectrum in the kination epoch is enhanced for both the cases over that in the radiation epoch. The dots show the best current 95$\%$ C.L. limits from different PTA experiments at frequency $f= 1 yr^{-1}$. Brown, orange and black dots represent the limits from EPTA, NANOGRAV and PPTA experiments respectively~\cite{Lentati:2015qwp,Arzoumanian:2015liz,Lasky:2015lej}.}  
\label{fig:2ndGW1}
}
\end{figure}
If the modes enter the horizon during RD epoch, they will evolve differently and second order transfer function for these modes is derived in the Ref.~\cite{Baumann:2007zm}. Combining the expression for $t(k, \eta)$ derived in this paper and in~\cite{Baumann:2007zm}, we get the full second order tensor transfer function for all the modes:

\begin{equation}
\label{equ:Finalt}
\hspace{-0.25cm}
t(k, \eta) = \left\{ \begin{array}{l l} 1 & \ \quad k \leq k_{\rm eq} \\  \Bigl( 
\frac{k}{k_{\rm eq}}\Bigr)^{-\gamma_2(k)} & \ \quad k_{\rm eq} \leq k < k_c(\eta) \\  
\left(\frac{k_{\rm T1}}{k}\right)^{\gamma_1(k)/2} \left(\frac{k_{\rm eq}}{k_{\rm T1}}\right)^{\gamma_2(k)}\   & \ \quad k_{\rm T1} \leq k < k_c(\eta)
\\
\frac{k_{\rm eq}}{k} \,\frac{a_{\rm eq}}{a(\eta)}  & \ \quad k_{\rm eq}<k<k_{\rm T1},~~k > k_c(\eta)\\
\left(\frac{k_{\rm T1}}{k}\right)^{1/2} \frac{k_{\rm eq}}{k_{\rm T1}} \,\frac{a_{\rm eq}}{a(\eta)}   & \ \quad k\geq k_{\rm T1},~~k > k_c(\eta)
\end{array} \right.
\end{equation}

In this work we have considered $T_1=10$MeV and $T_1=490$ GeV for computing the GW spectrum for two different peak scales of the primordial scalar power spectra. Using the asymptotic values of $\gamma_1(k)$ and $\gamma_2(k)$, for the modes that entered during SD era, we have calculated the value of critical scale at present time $\eta_0$ for these two different temperatures. Since the frequency ($f$) and wavenumber ($k$) are related as $k=2 \pi f/c$, we can calculate the critical frequency corresponding to critical wavenumber as $f_c(\eta)= \frac{c k_c (\eta)}{2 \pi}$. Values of the critical wavenumbers and critical frequencies for the modes entered the horizon during SD epoch are given in Table~\ref{Table2}. 
It is evident from the Table~\ref{Table2} that $k \geq k_{T1}>k_c(\eta_0)$ for the modes which  entered the horizon during SD epoch.
Therefore, these modes have not settled till present epoch. Moreover, for the modes which entered the horizon during RD era, critical scale at present epoch $k_c(\eta_0)=0.120$ Mpc$^{-1}$ and corresponding critical frequency $f_c(\eta_0)= 1.862\times10^{-16}$ Hz. 

\begin{table}[]
\caption{Values of critical wavenumber and critical  frequency for two different $T_1$ for the modes which has entered the horizon during SD era.}
\begin{center}

\resizebox{0.48\textwidth}{!}{
\begin{tabular}{|c|c|c|c|}
\hline 
$T_1 $ & $k_{T1} ({~\rm Mpc^{-1}})$ & $k_{c}(\eta_0) ({~\rm Mpc^{-1}})$ & $f_{c}(\eta_0) ({~\rm Hz})$\\
\hline
\hline 
$490 \rm{GeV} $ & $1.124\times10^{10}$ & 0.695 & $1.075\times10^{-15}$ \\
\hline
$ 10\rm{MeV} $ & $1.564\times10^{5}$ & 4.266 &  $6.597\times10^{-15}$\\
\hline
\end{tabular}}
\label{Table2}
\end{center}
\end{table}
Hence, for the modes with $k>k_c(\eta_0)$ or equivalently $f> f_c(\eta_0)$, second order tensor transfer function at present time $t(k, \eta_0)$, given by the last two expressions in Eq.~\eqref{equ:Finalt},
 is explicitly independent of both $\gamma_1(k)$ and $\gamma_2(k)$.
 In this manuscript we mainly are focusing on the modes which enter the horizon during SD epochs ($k\geq k_{\rm T1}$) and for these modes $k>k_c(\eta_0)$. Therefore, mathematically speaking, due to no explicit dependence of the transfer function on $\gamma_1(k)$ and $\gamma_2(k)$,  the analytical method in this manuscript successfully gives the correct result for the modes entering the horizon in SD epoch, as long as $k_{T1}>k_c(\eta _0)$. However, since we are using some approximations to calculate the values of $\gamma_1(k)$ and $\gamma_2(k)$, this is not an exact method to calculate $k_c$ and the GW spectrum for the modes close to $k_c$. Furthermore, If one wants to calculate the full GW spectrum for all the modes entering the horizon at any epoch, semi analytical approach, as followed in~\cite{Kohri:2018awv,Espinosa:2018eve}, has to be used to calculate the GW spectrum.

We now have all the necessary quantities to calculate the initial tensor power spectrum as well as the final GW spectrum. The fraction of second order GW energy density today, $\Omega^{(2)}_{\rm GW,\, 0} $ is then given by
\begin{eqnarray}
\Omega^{(2)}_{\rm GW,\,0}\,&=&\,\frac{a_0 k^2}{6 \pi^{2}a_{\rm eq}k_{\rm eq}^2}t^2(k,\eta_0) {P}^{(i)}_h (k,\eta_{\rm i}(k))\,.
\end{eqnarray}

In this paper we calculate the GW spectrum for two different primordial scalar power spectra. Furthermore, for each primordial scalar power spectrum, we have calculated the GW spectrum for two different values of peak scale $k_p$. In the first analysis, we have fixed $k_p=6\times10^{12}$ for which the scales with large primordial curvature fluctuations enter the horizon only during SD epoch. In this analysis we have used the first order scalar transfer function as given by Eq.~\eqref{equ:transfer}.

In the second analysis we have chosen $k_p=2\times10^6~{\rm Mpc^{-1}}$  and in this case the modes with large scalar fluctuations enter the horizon in both SD and RD epochs. Therefore, we will be considering $\eta_{\rm{eq}}>\eta\geq\eta_1$ regime also in this analysis. Hence, in the analysis with $k_p=2\times10^6~{\rm Mpc^{-1}}$, first order scalar transfer function will be different for modes entering the horizon during the SD and RD epochs (explained in detail in Appendix~\ref{sec:sca_trans}). We have used the full expression given by Eq.~\eqref{equ:Phiw} to calculate the GW spectrum in this analysis. 
These two values of $k_p$ are considered with $T_1=490~ {\rm GeV}$ (frequencies at the maximum sensitivity of LISA) and $T_1=10~ {\rm MeV}$ (close to the PTA relevant frequency) respectively. 

Now using the expression for $\Phi(k \eta)$ from Eq.~\eqref{equ:transfer} valid for the modes entering in SD epochs, we can rewrite the expression for $P^{(i)}_h (k)$ as
\begin{widetext}
\begin{eqnarray}
P^{(i)}_{h}(k,\eta_{\rm i}(k))\, \simeq  \frac{1}{\pi k} \int_0^{\infty} {\rm d }p \int_{-1}^1 {\rm d} \mu \,  \frac{ p^3 \left(1- \mu^2 \right)^2  P(p) P(\sqrt{|k^2+p^2 - 2\mu k p|}) }{\left( 1 + \bigg(\frac{p}{k}\bigg)^{3/2} \right)^2 \left( 1+ \bigg(\frac{|k^2 + p^2 - 2 \mu k p|^{1/2}}{k}\bigg)^{3/2} \right)^2 \left( |k^2+p^2 - 2\mu k p | \right)^{3/2}}~.
\label{equ:in_pow}
\end{eqnarray}
\end{widetext}
\subsubsection{Second order GW spectrum for Gaussian power spectrum}
Second order GW spectrum depends on the primordial scalar power spectrum. In our first analysis, we use the Gaussian power spectrum for curvature perturbations as given by Eq.~\eqref{pow3} to calculate the tensor power spectrum at horizon crossing. As mentioned in the previous section, we have calculated GW spectrum for two different values of peak scales. Values of all other parameters used in this analysis are given in Table~\ref{Table1} and Sec.~\ref{analysisPBH}. We have numerically solved the integral given by Eq.~\eqref{equ:phi0} to get the second order tensor power spectrum at horizon crossing. We also find that all the modes relevant to this analysis are greater than the critical scale ($k_c(\eta _0)$). Finally, we calculate the fraction of GW energy density in the Universe at present epoch $\Omega^{(2)}_{\rm GW,\,0}$ for both the peak scales.
In the analysis with peak scale $k_p=6\times10^{12}$ Mpc$^{-1}$, the GW spectrum peaks at $\tilde{k}_p\approx 6\times 10^{14}$ Mpc$^{-1}$ and the peak value of fraction of GW energy density in the Universe is $\approx7\times 10^{-9}$ . Whereas, in the analysis with $k_p=2\times10^{6}$ Mpc$^{-1}$, Gw spectrum peaks at $\tilde{k}_p\approx 2\times 10^{8}$ Mpc$^{-1}$ and its peak value is $\approx2.0\times 10^{-9}$ in the analysis with $k_p=2\times10^{6}$ Mpc$^{-1}$. The results for the GW spectra are shown in the Fig.s~\ref{fig:2ndGW} and \ref{fig:2ndGW1}. It is clear from the these figures that  if there is an additional SD era before the onset of RD epoch, then there will be a significant enhancement in the $\Omega^{(2)}_{\rm GW,\,0}$  as compared to that in a pure RD era. However, the form of the GW spectra in both the SD and standard RD era are similar for the Gaussian power spectrum.

\subsubsection{Second order GW spectrum for broken power law power spectrum}
Now, we calculate the GW spectrum for the broken power law scalar power spectrum of Eq.~\eqref{pow2}. Using this expression in Eq.~\eqref{equ:in_pow} and putting $x=p/k$, we get 
\begin{widetext}
\begin{equation}
\begin{aligned}
P^{(i)}_{h}(k,\eta_{\rm i}(k)) \,\simeq \,& \frac{2\mathcal{P}_{\rm p}^2}{\pi} \left(\frac{k}{k_{\rm p}}\right)^{2m} \int_0^{k_{\rm p}/k} {\rm d} x \int_{-1}^1 {\rm d}\mu \frac{(1-\mu^2)^2(1+x^2-2x\mu)^{\frac{m-3}{2}}x^{m+3}}{(1+(\sqrt{1+x^2-2x\mu})^{3/2})^2(1+x^{3/2})^2}\\
& + \frac{2\mathcal{P}_{\rm p}^2}{\pi} \left(\frac{k}{k_{\rm p}}\right)^{-2n} \int_{k_{\rm p}/k}^{\infty} {\rm d} x \int^{1}_{-1} {\rm d} \mu \frac{(1-\mu^2)^2(1+x^2-2x\mu)^{\frac{-n-3}{2}}x^{-n+3}}{(1+(\sqrt{1+x^2-2x\mu})^{3/2})^2(1+x^{3/2})^2}.
\end{aligned}
\end{equation}
\end{widetext}
Here also, we have used the two same values of peak scales as used in the case of analysis with Gaussian power spectrum. 
Values of all other parameters used in this analysis are given in Table~\ref{Table1} and in Sec.~\ref{analysisPBH}. Then we numerically solve the integral for tensor power spectrum at horizon crossing and calculate the GW spectrum. $\Omega^{(2)}_{\rm GW,\, 0} $ is plotted against frequency in Fig.~\ref{fig:2ndGW} and Fig.~\ref{fig:2ndGW1}. Again, it is evident from these figures that it reaches the maximum value at $\tilde{k}_p\approx k_p=6\times10^{12}$ Mpc$^{-1}$ and $\tilde{k}_p\approx k_p=2\times10^{6}$ Mpc$^{-1}$ for the two respective cases and after that becomes constant for both the cases. On the other hand $\Omega^{(2)}_{\rm GW,\, 0}$ starts to decrease after reaching the maximum value if there is no additional SD epoch. Therefore, there is an enhancement in GW spectrum obtained for an additional SD era as compared to that in the standard RD era. Moreover, the profile of the GW spectrum is different for SD era than for a pure RD epoch.

\section{Discussion and summary}
\label{discussion}
PBHs formed in the early universe can be considered as a probe for the smaller scales of inflation since the profile of the density contrast at postinflationary epochs depend on the primordial curvature power spectrum. In this paper, we argue that since the resulting abundance of PBHs depend also on the evolution of the background at the time of formation, hence alternate cosmological histories will lead to modification in the PBH abundance.

In this paper, given a primordial power spectrum with a rise in amplitude for the smaller scales, we have discussed the effect of an alternate pre-BBN cosmological history on the abundance of PBH and on the amplitude of first and second order GW spectra. Sec.~\ref{pbhform} discusses the modification of the PBH mass spectrum $f_{\rm PBH}(M)$ in an arbitrary pre-BBN SD epoch and comments on the possible gain in abundance in contrast with PBH formed in a radiation era. The analysis in Sec.~\ref{analysisPBH} explains that the requirement of the amplitude of primordial curvature power spectrum $P_{\zeta}(k)$ is lower for PBH formation in a SD epoch with $1/3<w\leq 1$ than in a radiation epoch. Exact parameter values for $f_{\rm PBH}^{\rm tot}\sim 10\%$ are quoted in Table~\ref{Table1}, which show that abundant PBH production in a kination epoch ($w=1$) would require much lower value of $P_p=P_{\zeta}(k_p)$. In the most optimistic scenario with $k_p=6\times 10^{12}~{\rm Mpc^{-1}}$ and a Gaussian primordial power spectrum, $P_p$ can be lowered by almost an order of magnitude with respect to RD for reaching the same abundance. Exact mass spectra are shown for two different $P_{\zeta}(k)$ in Fig.~\ref{fig_solar} and Fig.~\ref{fig_LISA} for two different positions of the peak at $k_p=2\times 10^{6}~{\rm Mpc^{-1}}$ and $k_p=6\times 10^{12}~{\rm Mpc^{-1}}$ respectively.

In Sec.~\ref{modGW}, we have analysed the effect of such an early SD epoch on the resulting GW spectrum. The first order GW spectrum calculated in subsection~\ref{modGW1} has an enhancement for the modes that enter the horizon during SD epoch due to modification in the transfer function in this epoch. The final $\Omega_{\rm GW,0}^{(1)}(k)$ depends on $w$, $T_1$ and $H_{\rm inf}$. Here, for an early kination epoch $w=1$ we have considered a particular frequency $f(T_1)\simeq 10^{-4}$ where this epoch ends and radiation starts to dominate at $T_1=490~{\rm GeV}$. In Sec.~\ref{modGW2}, we derive the second order GW spectra induced from scalar fluctuations.

The $T_1$ considered to calculate GW spectra in Sec.~\ref{modGW} is different than that considered while calculating PBH abundance in Sec.~\ref{analysisPBH}, for the reason that moving $T_1$ towards $T_{\rm BBN}$ enhances the GW power spectra even more which can overshoot the bound from CMB (magenta dashed line in Fig.~\ref{GW_order1}, Fig.~\ref{fig:2ndGW} and Fig.~\ref{fig:2ndGW1}). However, a high value of $T_1$ leads to a smaller gain in the GW spectra for the kination epoch as compared to radiation epoch. Then, to reach desirable PBH abundance, the power requirement may be a little higher than what is quoted in Table~\ref{Table1} for $w=1$, nevertheless, it will still be smaller than that required for $w=1/3$ since $\beta(M) > \beta ^{\rm rad}(M)$ in Eq.~\eqref{gain01}. 


An immediate consequence of including an additional epoch in the early universe is the modification in the matching equation~\cite{Adshead:2010mc}. If a mode $k=a_kH_k$ exited the inflationary horizon $N(k)$ e-folds before the end of inflation then
\begin{eqnarray}
e^{N(k)}&=& \frac{a_{\rm end}}{a_k}\nonumber \\
        &=& \frac{H_k}{H_{\rm end}}\frac{a_{\rm end}H_{\rm end}}{a_{\rm reh}H_{\rm reh}}\frac{a_{\rm reh}H_{\rm reh}}{a_0H_0}\frac{a_0H_0}{k},\label{mathcing1}
\end{eqnarray}
where $a_{\rm end}$, $a_{\rm reh}$ and $a_0$ are the scale factors at the end of inflation, at the end of reheating and at present respectively. $H_{\rm end}$, $H_{\rm reh}$ and $H_0$ are the Hubble parameters at the end of inflation, at the end of reheating and at present respectively.
But, with instant reheating ($a_{\rm end} = a_{\rm reh}$) and an additional epoch of kinetic domination $w=1$, as considered in the present manuscript, Eq.~\eqref{mathcing1} is modified as:
\begin{equation}
e^{N(k)}=\frac{H_k}{H_{\rm end}}\frac{a_{\rm end}H_{\rm end}}{a (T_1)H(T_1)}\frac{a (T_1)H (T_1)}{a_0H_0}\frac{a_0H_0}{k},\label{matchingkine}
\end{equation}
where $\frac{a(T_1)}{a_{\rm end}}=\big(\frac{\rho _{\rm end}}{\rho (T_1)}\big)^{1/6}$. After some algebraic manipulations, we reach at:
\begin{eqnarray}
N(k)&\simeq 56.12 - \log \frac{k}{k_*}+\frac{1}{6}\log \frac{2}{3}+\log \frac{V_k^{1/4}}{V_{\rm end}^{1/4}}\nonumber \\
&-\frac{1}{3}\log \frac{\rho (T_1)^{1/4}}{V_{\rm end}^{1/4}}+\log \frac{V_k^{1/4}}{10^{16}{\rm GeV}}.\label{match2}
\end{eqnarray}
Therefore, given a model of inflation (i.e. $V_k$ and $V_{\rm end}$), $T_1$ can be written as a function of $N(k_*)$ using Eq.~\eqref{rho1T1} and Eq.~\eqref{match2}. If we consider slow roll inflation with $V_{k_*}\simeq V_{\rm end} = 10^{16}GeV$ then
\begin{equation}
T_1=e^{3(56.12-N(k_*))}\times \bigg(\frac{2}{3}\bigg)^{1/2}\bigg(\frac{\pi^2g_*(T_1)}{15}\bigg) ^ {-1/4}.\label{T1Nk}
\end{equation}
In this paper, we have considered $T_1=10{~\rm MeV}$ while calculating PBH abundance and GW spectra with $k_p=2\times10^6~{\rm Mpc^{-1}}$ which corresponds to $N(k_*)=67.5$. We take $T_1=490{~\rm GeV}$ while calculating GW spectra with $k_p=6\times10^{12}~{\rm Mpc^{-1}}$, which corresponds to $N(k_*)=66.2$. For an attractor type inflation potential with $n_s\simeq 1-2/N(k_*)$, these two values of $T_1$ predict $n_s$ values that are within the allowed range $n_s=0.9649\pm 0.0042$ ($68\%$ C.L.) from Planck 2018~\cite{Akrami:2018odb}. 

As discussed in Sec.~\ref{pbhform}, the exact dependence of the critical density contrast $\delta_c$ on the background e.o.s. $w$ is not well understood. In earlier papers~\cite{Carr:1975qj} $\delta_c =w$ was considered with a discussion of uncertainties of order unity in the numerical prefactor of $w$. Subsequent full numerical analysis has put the critical value for RD to be $\delta_c^{w=1/3}=0.67-0.71$ in~\cite{Niemeyer:1999ak}, whereas considering a pure growing mode,~\cite{Musco:2004ak} reported $\delta_c^{w=1/3}=0.43–0.47$. The exact relation that is used here is Eq.~\ref{deltac_w} (calculated in~\cite{Harada:2013epa}) which predicts $\delta_c^{w=1/3}\simeq 0.414$. The uncertainty in $\delta_c$ can lead to a large uncertainty in the PBH abundance due to the exponential dependence of the mass fraction $\beta (M)$ on $\delta_c$ (Eq.~\eqref{betadef}). However, using Eq.~\eqref{deltac_w}, $\delta_c$ decreases from this value as $w$ increases from $1/3$, which helps here to reach a higher abundance of PBH for the same amplitude of primordial density fluctuations.

The PBH analysis in Sec.~\ref{pbhform} can also be done using Peak theory (PT) method instead of Press-Schechter (PS) formalism. PT formalism focuses on the peak profile of either the metric perturbation $\zeta (\hat{r})$, or the curvature perturbation $K(r)$~\cite{Musco:2018rwt,Kalaja:2019uju,Germani:2018jgr}. In terms of $K(r)$, the perturbed metric is:
\begin{equation}
ds^2=-dt^2+a^2(t)\bigg(\frac{dr^2}{1-K(r)r^2}+r^2d\Omega ^2\bigg).
\end{equation}
The averaged density contrast, which is  a more relevant quantity of interest in case of an extended peak profile of $K(r)$, can be written (at horizon crossing) as:
\begin{equation}
\tilde{\delta } (r)=f(w)K(r) r^2, \label{tildel1}
\end{equation}
where $r$ is the radius of the spherical comoving volume on which it has been averaged. The coordinate origin is at the location of the peak. PBH formation criteria is expressed in terms of the compaction function, which is defined as the ratio of the mass excess over the physical radius: $\mathcal{C}(r,t)=\frac{2\delta M (r,t)}{R(r,t)}$.
Now, for a particular peak profile of $K(r)$ or $\zeta (\hat{r})$, there are two scales of importance: the scale $r_0$ where the local density contrast crosses zero and the scale $r_m$ where the compaction function reaches maximum value. Thus, $\tilde{\delta }_0 (r)=f(w)K(r) r_0^2$ and $\tilde{\delta }_m (r)=f(w)K(r) r_m^2$ with $f(w)=\frac{3(1+w)}{(5+3w)}$. The $\delta$ considered in a PS formalism is equivalent to $\tilde{\delta }_0$, but $\tilde{\delta }_m$ and $\tilde{\delta }_0$ are different in general. 

For a particular peak profile for curvature, one can find out $r_0$, $r_m$, $\tilde{\delta }_0$ and $\tilde{\delta }_m$ in terms of the profile parameters. Then, knowing the critical value of $\tilde{\delta }_0$, we can find the critical value for $\tilde{\delta }_m$. For a Gaussian curvature profile:
\begin{equation}
K(r)=\mathcal{A}\exp\bigg(-\frac{r^2}{2\Delta ^2}\bigg),\label{prof1}
\end{equation}
our numerical formula for $\tilde{\delta }_0^c$ in Eq.~\eqref{deltac_w} gives:
\begin{equation}
\tilde{\delta }_m^c=\frac{2e^{1/2}}{3}\tilde{\delta }_0^c=\frac{2e^{1/2}}{3} f(w)\sin ^2\bigg(\frac{\pi \sqrt{w}}{1+3w}\bigg).\label{deltacm1}
\end{equation}
The RD values are $\tilde{\delta }_0^{c,{\rm RD}}=0.414$ and $\tilde{\delta }_m^{c,{\rm RD}}=0.455$, whereas the $w=1$ epoch has $\tilde{\delta }_0^{c,w=1}=0.375$ and $\tilde{\delta }_m^{c,w=1}=0.412$. Thus, PT calculations also yield $\tilde{\delta }_m^{c,w=1}<\tilde{\delta }_m^{c,{\rm RD}}$, which is key to have positive gain in PBH abundance in the $w=1$ epoch.

\begin{figure}[]
\begin{center}
\includegraphics[width=0.45\textwidth]{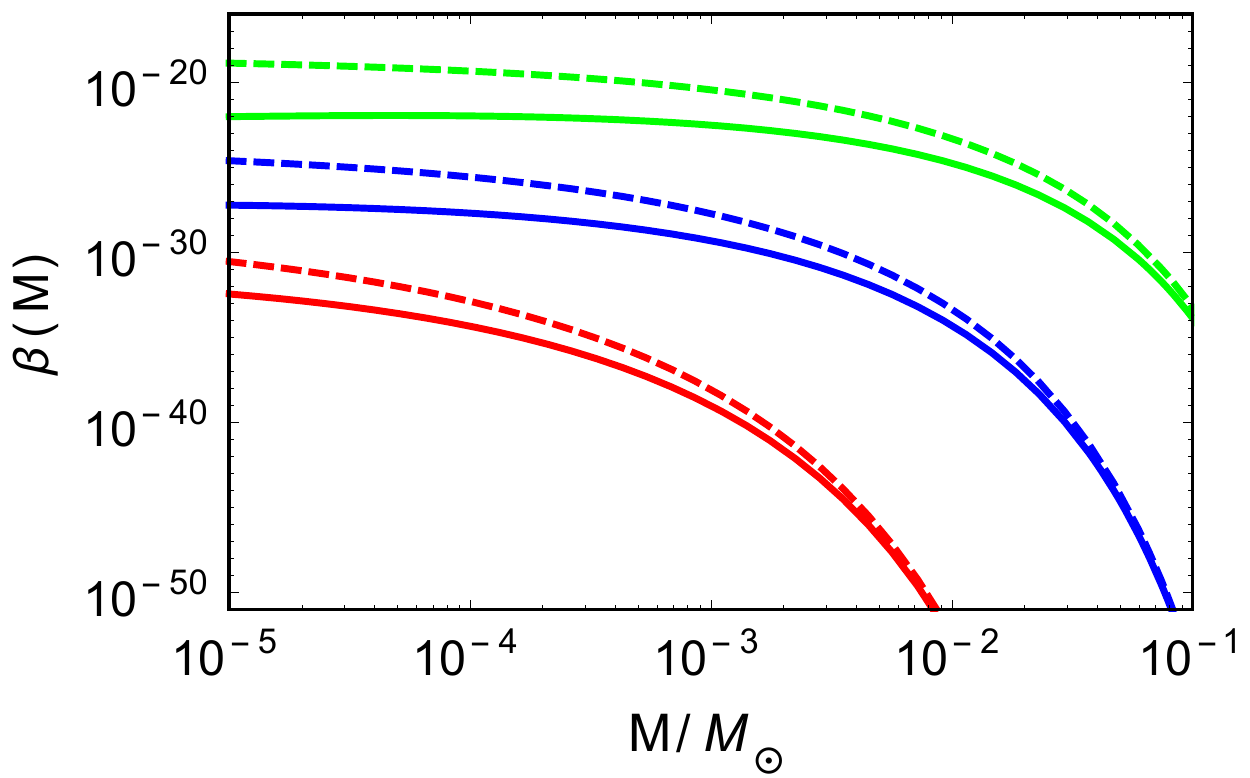}
\caption{PBH mass fraction $\beta(M)^{\rm PS}$ calculated in Press-Schechter formalism in dashed lines and $\beta(M)^{\rm PT}$ calculated in peak theory method in solid lines for $w=1/3$ (RD, red), $w=2/3$ (blue) and $w=1$ (green).}
\label{beta_PSPT}
\end{center}
\end{figure}

The PBH abundance can also be calculated explicitly using PT for the RD and $w=1$ case. As an example, following~\cite{Green:2004wb}, we plot in Fig.~\ref{beta_PSPT} the mass fraction $\beta (M)$ for the broken power law power spectrum given in Eq.~\eqref{pow2} with $k_p=2\times 10^6$ Mpc$^{-1}$ for both PS and PT formalism.
We note that there is an evident difference for the PS and PT calculations of PBH abundance for a particular $w$, which is expected as pointed out in Fig.1 of~\cite{Green:2004wb}. However, whether the formalism is PS or PT, increasing $w$ over $1/3$ increases the abundance by several orders of magnitude in both of these cases. Thus, the PT calculations for $\tilde{\delta }_m^c$ and $\beta(M)^{\rm PT}$  show that our goal to obtain larger abundance owing to a smaller value of critical density is fulfilled for the $w=1$ case as compared to RD. Therefore, PS and PT formalisms are consistent in fulfilling the motivation of this work.

Present and future GW interferometers such as aLIGO and LISA can constrain the scale of inflation. Fig.~\ref{GW_order1} shows that nonobservation of primordial GW signal in LIGO O1 and O2 constrain $H_{\rm inf}\leq 10^{12}~{\rm GeV}$, which is consistent with the current upper bound on the tensor-to-scalar ratio $r<0.067$ from Planck~\cite{Akrami:2018odb}. However, this result is for $T_1=490~{\rm GeV}$ and if the SD epoch sustains for a shorter time until a higher value of $T_1$, then the resulting $\Omega_{\rm GW,0}^{(1)}(k)$ will be smaller and therefore a higher value of $H_{\rm inf}$ can still be allowed by present LIGO observations.

Large primordial curvature perturbations at small scales that are relevant for PBH formation can source second order GW. In case of an early SD epoch, the second order GW profile is calculated in subsection~\ref{modGW2} for the broken power law and Gaussian primordial spectrum. The frequency $f\sim 0.01~{\rm Hz}$ relevant to the future sensitivity range of LISA corresponding to $k_p=6\times 10^{12}~{\rm Mpc^{-1}}$. The analysis for  $f\sim 10^{-9}~{\rm Hz}$ where PTAs are probing which corresponds to $k_p=2\times 10^{6}~{\rm Mpc^{-1}}$. We found that for the same primordial spectra, the second order GW amplitude is enhanced in a kination epoch as compared to a pure radiation epoch. It is evident from Fig.~\ref{fig:2ndGW1} that the models for the primordial curvature power spectra considered here are in tension with the current PTA observations for $k_p=2\times 10^{6}~{\rm Mpc^{-1}}$. However, varying the model parameters, specifically, decreasing $P_p$ would still allow these power spectra to conform with the PTA data. Upcoming LISA observation will be crucial to constrain the equation of state of a possible early SD epoch given the small scale primordial spectrum shows features near $k\sim 10^{12}~{\rm Mpc^{-1}}$.

Any observed GW spectrum in LISA can have contributions from all orders in perturbation theory whereas contributions from second and higher orders depend on the primordial curvature perturbations. In our case, the orders of the first and second order GW spectrum are quite different considering $H_{\rm inf}\sim 10^{12}~{\rm GeV}$ and the second order contribution is dominant, which can be probed by LISA. But degeneracies in the amplitude of GW spectrum originating at different orders of perturbation theory may arise for other cases with a different $P_{\zeta}(k)$ or different choices of $w$ and $T_1$. A possible way to break the degeneracy in the contribution from first and second order is that the first order GW spectrum does not depend on the feature of $P_{\zeta}(k)$, whereas the second order GW spectrum tracks the profile of $P_{\zeta}(k)$.

On the same note, we emphasize here that the second order GW spectrum depends both on the primordial curvature spectrum from inflation and on the evolution of the postinflationary background. Therefore, an independent observation of (stochastic) GW spectrum, even though confirmed to result from second order perturbation theory, is not sufficient to comment about the primordial inflationary dynamics at small scales, and this is one of the main findings of our analysis. This is evident from the second order GW spectra for the case of broken power-law power spectrum in Fig.~\ref{fig:2ndGW} and Fig.~\ref{fig:2ndGW1}. Despite originating from the same primordial curvature spectrum $P_{\zeta}(k)$ in Eq.~\eqref{pow2}, the GW spectra in Fig.~\ref{fig:2ndGW} for pure radiation domination (dashed blue line) and an early kination epoch (solid blue line) have very different profiles. Moreover, for kination epoch, the high-frequency arm of the GW spectrum attains a constant value because of (i) the modified form of the second order tensor transfer functions (last expression in Eq.~\eqref{equ:Finalt}) and (ii) the choice of $n$ in Eq.~\eqref{pow2}. With this choice, the constant value of $\Omega_{\rm GW,0}^{(2)}(k)$ can be extended until $f_{\rm max}=k_{\rm max}/2\pi$, where $k_{\rm max}=a_{\rm end}H_{\rm end}$ corresponds to the mode that went out of the horizon at the end of inflation. Therefore, given a particular model of inflation producing a broken power law power spectrum like Eq.~\eqref{pow2}, aLIGO and LISA observations can constrain the model in presence of a postinflationary SD epoch.

The effects of an early kination epoch on PBH formation and GW discussed in this work points to an interesting path to phenomenology by simultaneously considering the observational constraints on the PBH mass spectrum and observational constraints on the amplitude and feature of the GW spectrum using present and future GW surveys. The analysis for GW focuses on the proposed sensitivities and relevant frequency ranges for future LISA mission and present and future PTA surveys. With the proposed improvement in precision for the astrophysical and cosmological experiments constraining PBH abundance and with the proposed high sensitivity of future GW missions such as LISA and aLIGO, DECIGO and PTA, a rich phenomenological understanding of the pre-BBN epoch is expected and intended. It will be exciting to find a probe to distinguish between the contributions to the second order GW spectrum from inflationary power spectra and from postinflationary transfer functions. Moreover, the theories leading to a nonstandard pre-BBN epoch, such as quintessential inflation leading to an early kination epoch, can be constrained with such a phenomenological analysis. 

The exact implications of the inflationary power spectrum for PBH formation in a kination epoch can be interpreted in terms of parameters and field values of the underlying model of inflation. For a deviation from the standard hot big bang evolution with an additional early SD epoch, the respective field values and parameter ranges for inflation will be different for abundant PBH formation, which may be checked with the theoretically allowed parameter ranges for the particular model.

An early SD epoch, particularly a pre-BBN kination epoch may also have interesting predictions for the spin of the PBH formed in this epoch {due to higher pressure in the background~\cite{He:2019cdb,Mirbabayi:2019uph,DeLuca:2019buf}}. The merger rate for PBH formed in such an epoch~\cite{Korol:2019jud} can also be calculated. We hope to explore these aspects of nonstandard pre-BBN cosmology in the future.

\section*{Acknowledgement}
The authors sincerely thank the referees for their insightful comments and suggestions that helped in improving the structure and contents of this paper. SB is supported by a postdoctoral fellowship from Physical Research Laboratory, India. SB sincerely thanks A. De Felice, A. Di Marco and S. P. Patil for useful discussions about this work.

\newpage
 
\appendix

\section{Transfer function for first-order scalar modes}\label{sec:sca_trans}
In the absence of entropy perturbation, evolution of $\Phi_{\mathbf{p}}$ in any epoch with equation of state parameter $w$, is governed by the following equation (see Eq.(B3) in~\cite{Baumann:2007zm}).

\begin{equation}
\label{equ:Bardeen}
\Phi_{\mathbf{p}}'' + \frac{6(1+w)}{1+3w} \frac{1}{\eta} \Phi_{\mathbf{p}}' + w 
p^2 \Phi_{\mathbf{p}} = 0\, .
\end{equation}
The exact solution for this equation can be obtained in terms of Bessel functions, which is given as 
\begin{equation}
\label{sol:Bardeen}
\Phi_{\mathbf{p}}(\eta) = z^{-\alpha} \Bigl[ C_1(p) J_\alpha(y) + C_2(p) 
Y_\alpha(y)\Bigr]\, .
\end{equation}
Where $z \equiv \sqrt{w} p \eta$, $\alpha 
\equiv \frac{1}{2} \left(\frac{5+3 w}{1 + 3w} \right)$ and $J_\alpha$ and $Y_\alpha$ are Bessel functions of order $\alpha$. 
During the kination epoch ($w=1$), Eq.~\eqref{sol:Bardeen} takes the following form
\begin{equation}
\Phi_{\mathbf{p}}(\eta) =\frac{1}{p \eta}  C_1(p) J_1(p \eta) \, ,~~~~~~\eta < \eta_1
\end{equation}
where we have dropped the second term since it is the decaying solution. This expression gives the primordial value of $\Phi_\mathbf{p}$ in the early time limit $z =  p \eta \ll 1$ (superhorizon limit)
\begin{equation}
\lim_{y \to 0} \Phi_{\mathbf{p}}(\eta) = \frac{C_1(p)}{2} = \psi_{\mathbf{p}}\, .
\end{equation}
Hence we can write
\begin{equation}
\Phi_{\mathbf{p}}(\eta) = \psi_{\mathbf{p}} \Phi(p \eta)  \, ,~~~~~~\eta < \eta_1
\end{equation}
where $\Phi(p \eta)=\frac{2}{p \eta} J_1(p \eta) $ is the first order scalar transfer function during kination epoch.
Now for the superhorizon modes ($p \eta \ll 1$) during SD epoch,
\begin{equation}
\Phi(p \eta) = 1 + {\cal O}((p \eta)^2)\, , \quad \eta < \eta_{1}\,
\end{equation}
and for the subhorizon modes ($p \eta > 1$),
\begin{equation}
\Phi(p \eta)
\simeq \frac{1}{( p \eta)^{3/2}}\cos\left(\frac{3\pi}{2} -p \eta\right) . \quad \eta < \eta_{1}\,
\end{equation}
This can be written as following, which will be valid for both super- and subhorizon modes during the SD epoch.
\begin{equation}
\label{equ:transfer}
\Phi(p \eta) = \frac{1}{1+ (p \eta)^{3/2}}\, , \quad \eta < \eta_{1}\, 
\end{equation}
where we have neglected the oscillation term.

Now the solution to Eq.~\eqref{equ:Bardeen} during the RD epoch is
\begin{equation}
\Phi_{\mathbf{p}}(\eta) =\frac{3\sqrt{2}}{p \eta \pi} \Bigl[ C_3(p) j_{1}\left(\frac{p \eta}{\sqrt{3}}\right) + C_4(p) 
y_{1}\left(\frac{p \eta}{\sqrt{3}}\right)\Bigr]\,,~\eta_1 \leq \eta < \eta_{\rm{eq}}
\end{equation}
where $j_{1}\left(\frac{p \eta}{\sqrt{3}}\right)$ and $y_{1}\left(\frac{p \eta}{\sqrt{3}}\right)$ are the spherical Bessel function of first and second type.
\begin{equation}
j_{1}\left(\frac{p \eta}{\sqrt{3}}\right)= \frac{\sin(p \eta/\sqrt{3})}{(p \eta/\sqrt{3})^2} - \frac{\cos(p \eta/\sqrt{3})}{p \eta/\sqrt{3}},
\end{equation}
\begin{equation}
y_{1}\left(\frac{p \eta}{\sqrt{3}}\right)= -\frac{\cos(p \eta/\sqrt{3})}{(p \eta/\sqrt{3})^2} - \frac{\sin(p \eta/\sqrt{3})}{p \eta/\sqrt{3}}.
\end{equation}
Coefficients $C_3(p)$ and $C_4(p)$ are calculated using the continuity condition at the junction of the SD and the RD epoch: $\Phi_{\mathbf{p}}^{\rm SD}(\eta_1)=\Phi_{\mathbf{p}}^{\rm RD}(\eta_1)$ and $\Phi_{\mathbf{p}}'^{\rm SD}(\eta_1)=\Phi_{\mathbf{p}}'^{\rm RD}(\eta_1)$. Using these conditions, $C_3(p)$ and $C_4(p)$ are solved to be
\begin{widetext}
\begin{equation}
C_3(p)=\sqrt{2}\pi\psi_{\mathbf{p}}\bigl[\frac{z_1\left(J_{1}(z_1)y_{0}(z_1)-J_{0}(z_1)y_{1}(z_1)+J_{2}(z_1)y_{1}(z_1)-J_{1}(z_1)y_{2}(z_1)\right)-J_{1}(z_1)y_{1}(z_1)}{z_1\left(j_{1}(z_1)y_{0}(z_1)-j_{0}(z_1)y_{1}(z_1)+j_{2}(z_1)y_{1}(z_1)-j_{1}(z_1)y_{2}(z_1)\right)}\bigr]
\end{equation}
\begin{equation}
C_4(p)=\sqrt{2}\pi\psi_{\mathbf{p}}\bigl[\frac{z_1\left(J_{1}(z_1)j_{0}(z_1)-J_{0}(z_1)j_{1}(z_1)+J_{2}(z_1)j_{1}(z_1)-J_{1}(z_1)j_{2}(z_1)\right)-J_{1}(z_1)j_{1}(z_1)}{z_1\left(j_{1}(z_1)y_{0}(z_1)-j_{0}(z_1)y_{1}(z_1)+j_{2}(z_1)y_{1}(z_1)-j_{1}(z_1)y_{2}(z_1)\right)}\bigr]\,
\end{equation}
\end{widetext}
where $z_1=\frac{p \eta_1}{\sqrt{3}}$.
Hence, the first order scalar transfer function during RD epoch is given as 
\begin{equation}
\label{eq:rdtrans_first}
\Phi(p \eta) = \frac{3\sqrt{2}}{p \eta \pi} \Bigl[ A(p) j_{1}\left(\frac{p \eta}{\sqrt{3}}\right) + B(p) 
y_{1}\left(\frac{p \eta}{\sqrt{3}}\right)\Bigr]\, ,~~\eta_1\leq\eta < \eta_{\rm{eq}}\, ,
\end{equation}
where $A(p)=\frac{C_3(p)}{\psi_{\mathbf{p}}}$ and $B(p)=\frac{C_4(p)}{\psi_{\mathbf{p}}}$. For the superhorizon modes (p $\eta\ll 1$) during RD epoch, we find that
$\Phi(p \eta) \approx 1$. Whereas, for the subhorizon modes (p$\eta > 1$) during RD epoch
\begin{equation}
\label{eq:rdtrans_second}
\Phi(p \eta) = -\frac{3\sqrt{6}}{(p \eta)^2 \pi} \Bigl[ A(p)\cos(\frac{p \eta}{\sqrt{3}}) + B(p)\sin(\frac{p \eta}{\sqrt{3}}) 
\Bigr]\, ,~~\eta_1 \leq \eta < \eta_{\rm{eq}}\, .
\end{equation}
Now, of all the modes that are subhorizon during the RD epoch, some have entered the horizon during the SD epoch and others cross the horizon during RD epoch itself.
For the modes which have entered the horizon during SD era ($p \geq p_{T1}$), the initial condition for their evolution in the RD epoch is different than their typical superhorizon freeze-in values and thus, the expression
in Eq.~\eqref{eq:rdtrans_second} for these modes is different from the expression derived in Ref~\cite{Baumann:2007zm} for the case of standard postinflationary RD epoch.

Since the modes entering in the SD epoch correspond to $p > p_{T1}$, where $p_{T1}\sim\frac{1}{\eta_1}$, so that $p\eta_1 > 1$ for these modes.
Therefore,  in the limit $p \eta_1 > 1$ 
\begin{equation}
\label{ApBp}
A(p) \approx \frac{\sqrt{2 \pi} }{3^{1/4}} \, (p\eta_1)^{1/2} ~~~{\rm and} ~~B(p) \approx \frac{\sqrt{2 \pi} }{3^{1/4}} \, (p\eta_1)^{1/2}\, , 
\end{equation}
where we have used the asymptotic limit for the Bessel functions.
Using Eq.~\eqref{ApBp} in Eq.~\eqref{eq:rdtrans_second}, we get the first order scalar transfer function in RD era for the modes which entered the horizon in SD era ($p> p_{T1}$) as
\begin{equation}
\label{scsdrd}
\Phi(p \eta) \approx -\frac{3^{3/4}\sqrt{12 \pi}}{(p \eta)^2 \pi} \, (p\eta_1)^{1/2} \Bigl[ \cos(\frac{p \eta}{\sqrt{3}}) + \sin(\frac{p \eta}{\sqrt{3}}) 
\Bigr]\, ,~~\eta_1 \leq \eta < \eta_{\rm{eq}}\, .
\end{equation} 
However, for the modes which were superhorizon during SD era and are entering the horizon only during RD era ($p_{T1}\leq p \leq p_{eq}$), Eq.~\eqref{eq:rdtrans_second} goes back to the expression derived in ref~\cite{Baumann:2007zm}. Since $p_{T1}\sim \frac{1}{\eta_1}$ and $p_{T1}\geq p$, $p\eta_1< 1$ for these modes. Therefore, in the limit $p\eta_1< 1$, 
\begin{equation}
A(p) \approx \frac{(15+\sqrt{3})\pi}{6 \sqrt{2}}~~\rm{and} ~~ B(p)\approx 0\, .
\end{equation}
Hence the first order scalar transfer function during RD era for these modes ($p_{T1}\leq p \leq p_{eq}$) is given as
\begin{equation}
\label{eq:rdtrans_first}
\Phi(p \eta) = \frac{(15+\sqrt{3})}{2} \frac{1}{p \eta }  j_{1}\left(\frac{p \eta}{\sqrt{3}}\right)\, 
,~~~~\eta_1 \leq \eta < \eta_{\rm{eq}}\,.
\end{equation}
This is same as the expression of $\Phi(p \eta)$ derived in ref~\cite{Baumann:2007zm} for the case of standard postinflationary RD epoch.

\section{Exponents $\gamma _1$ and $\gamma _2$ in the transfer function for second order tensor modes}\label{sec:tens_trans}
Let $x \equiv |\mathbf{k - p}| \eta$ and $y \equiv |\mathbf{ p}| 
\eta$. 
Using these definitions of $x$ and $y$, we write the source term Eq.~\eqref{equ:source} as
\begin{equation}
\label{equ:s1}
{\cal S} (\mathbf{k}, \eta)
=  \frac{2 \pi}{\eta^5} \int_0^\infty y^4 d y \int_{-1}^1 d \mu\, [1-\mu^2]\, f(x,y)\, \psi_{\mathbf{k-p}} \psi_{\mathbf{p}}\, ,
\end{equation}
where
\begin{equation} 
\label{eq:fxy}
f(x,y) = \frac{8\big[ (5+3 w) \Phi(x) \Phi(y)  + 2 \left( 2 \Phi(x) + x \Phi^x(x) \right)
  y \Phi^y(y)\big]}{3(1+w)} .
\end{equation}
Here we have defined $\Phi^x(x)= \frac{\partial \Phi(x)}{\partial x}$ and $\Phi^y(y)= \frac{\partial \Phi(y)}{\partial y }$.
Now, the primordial scalar power is 
\begin{equation}
\label{equ:psip}
\psi_{\mathbf{k-p}} \psi_{\mathbf{p}} \propto 
\frac{\eta^3}{x^{3/2} y^{3/2}}\, .
\end{equation}
Using Eq.~\eqref{equ:psip} in Eq.~\eqref{equ:s1}, we get
\begin{equation}
\label{equ:int}
{\cal S} (\mathbf{k}, \eta)
\propto  \frac{2 \pi}{\eta^2} \int_0^\infty d y \int_{-1}^1 d \mu\, [1-\mu^2] 
\,
        \frac{y^{5/2}}{x(y,\mu)^{3/2}} \, f(x,y)\, .
\end{equation}
Let us first see the time evolution of Eq.~\eqref{equ:int} during SD era for the modes which entered the horizon during SD era only.
In the SD era, first order scalar transfer function,  $\Phi(x)=\frac{1}{1+ (x)^{3/2}}$ (Eq.~\eqref{equ:transfer}), therefore
\begin{equation}
\label{f:sd}
f(x,y)=\frac{1}{(1+x^{3/2})(1+y^{3/2})} \left[ \frac{32}{3} - \frac{2(4+x^{3/2})y^{3/2}}{(1+x^{3/2})(1+y^{3/2})} \right]\,.
\end{equation} 
Let us now estimate the integral in Eq.~\eqref{equ:int}
for the modes which entered the horizon during SD era. First, definitions of $x$ and $y$ lead to:
\begin{equation}
\label{eq:relpk}
\Bigl( \frac{x(y,\mu)}{k \eta} \Bigr)^2
= 1 + \Bigl( \frac{y}{k \eta} \Bigr)^2 - 2\Bigl( \frac{y}{k \eta} \Bigr) \mu\, .
\end{equation}
Now, in the limit $x \rightarrow 0\,(y\rightarrow k \eta\,,\,\mu\rightarrow 1)$, the integrand goes to zero. To see this, first take $y\rightarrow k \eta$ in Eq.~\eqref{eq:relpk}
\begin{equation}
\label{eq:x/keta}
\Bigl(\frac{x}{k \eta} \Bigr)^2 \to 2 [1-\mu]\,.
\end{equation}
Now, using Eq.~\eqref{eq:x/keta}, we calculate the $\frac{1-\mu^2}{x^{3/2}}$ in the limit $y\rightarrow k \eta$  and find 
\begin{equation}
\label{eq:x/keta2}
\frac{1-\mu^2}{x^{3/2}} \to \frac{(1-\mu)^{1/4} (1+\mu)}{(2 k\eta)^{3/4}}\, .
\end{equation}
It is clear from Eq.~\eqref{eq:x/keta2} that $\frac{1-\mu^2}{x^{3/2}}$ goes to zero as $\mu\rightarrow 1$, which implies that integrand in Eq.~\eqref{equ:int} goes to zero as $x \rightarrow 0$. Whereas, the integrand is suppressed by the factor $y^{5/2}$ in the limit $y  \rightarrow 0$. Since $\lim_{x\to\infty} f(x,y) =0$ and $\lim_{y\to\infty} f(x,y) =0$, the large x and y limit of the integrand is suppressed by $f(x,y)$. Therefore, the dominant contribution to the integral (\ref{equ:int})  comes from the
regions in the phase space where $\rm p \sim k$ (\textit{i.e.} $y \sim k \eta$) and 
$|\mathbf{k-p}| \sim k$ (\textit{i.e.} $x \sim k \eta$, $\mu \sim 0$).
Let us therefore rewrite the source term (\ref{equ:int}) using Eq.~\eqref{f:sd} as
\begin{eqnarray}
&{\cal S}&\, \propto \frac{1}{\eta^2} \int \rm d \ln y \int \rm d \ln (1-\mu) \, (1-\mu)^2 
(1+\mu) \frac{y^{7/2}}{x^{3/2}} \nonumber \\ &\,&\frac{1}{(1+x^{3/2})(1+y^{3/2})} \left[ \frac{32}{3} - \frac{2(4+x^{3/2})y^{3/2}}{(1+x^{3/2})(1+y^{3/2})} \right]\,. 
\end{eqnarray}
Now, take the limit $x,y \to k \eta > 1$ and $\mu \to 0$ to get
\begin{eqnarray}
{\cal S} &\propto& \frac{1}{\eta^2} \frac{(k\eta)^{7/2}}{(k \eta)^{3/2}} 
\frac{1}{[(k \eta)^{3/2} +1]^2} \left[16 -  3 \frac{(k \eta)^{3}}{[(k \eta)^{3/2} +1]^2} 
\right] \nonumber \\
& \approx &\frac{1}{\eta^2} \frac{1}{(k \eta)} \propto \frac{1}{a^6}\, .
\end{eqnarray}\label{eq:sca_trans}
where we have used the fact that scale factor $\eta\propto a^2$ during SD epoch. Therefore, for the modes which entered the horizon during SD era, source term can decay at most as $a^{-6}$ during SD era. This implies that $\gamma_1(k)\leq 6$  during SD era.

Let us now see the time evolution of source term during RD era ($\eta_{1}\leq \eta \leq \eta_{eq} $) for the modes which entered the horizon during SD era ($ k > k_{T1} $). For the modes with $ k > k_{T1} $, first order scalar transfer function during RD epoch is given by Eq.~\eqref{scsdrd}.
Using Eq.~\eqref{scsdrd} in Eq.~\eqref{eq:fxy}, we get
\begin{widetext}
\begin{equation}
\label{fxy:sdrd}
f(x,y)=\frac{12 \eta_1\left[\sqrt{3}\left((27+4 x y) \cos(\frac{x-y}{\sqrt{3}})+(27 - 4 x y) \sin(\frac{x+y}{\sqrt{3}})\right)+ 6 \left( (-3x+y) \cos(\frac{x+y}{\sqrt{3}})+(3x+y) \sin(\frac{x-y}{\sqrt{3}})\right)\right]}{\pi\, x^{3/2} y^{3/2} \eta}\,.
\end{equation}
\end{widetext}
We find that $\lim_{x\to\infty} f(x,y) =0$ and $\lim_{y\to\infty} f(x,y) =0$. Hence, Integrand in Eq.~\eqref{equ:int}  is suppressed for large value of x and y. Again, the dominant contribution to the integral (\ref{equ:int})  comes from 
regions where $\rm p \sim k$ (\textit{i.e.} $y \sim k \eta$) and 
$|\mathbf{k-p}| \sim k$ (\textit{i.e.} $x \sim k \eta$, $\mu \sim 0$). Therefore, with the above limits, the source term (\ref{equ:int}) using Eq.~\eqref{fxy:sdrd} is:
\begin{widetext}
\begin{equation}
{\cal S} \propto  \frac{1}{\eta^2} \frac{(k\eta)^{7/2}}{(k\eta)^{3/2}}\frac{12\eta _1}{\eta}\frac{\bigg[27\sqrt{3}\bigg(1+\sin(\frac{2k \eta}{\sqrt{3}})\bigg)+4\sqrt{3}(k\eta)^2\bigg(1-\sin(\frac{2k \eta}{\sqrt{3}})\bigg) - 12 (k\eta)\cos(\frac{2k \eta}{\sqrt{3}})\bigg]}{\pi (k\eta)^3}\, .
\end{equation}
\end{widetext}
In this expression, in the limit $ k \eta > 1$, the dominant contribution comes for the second term within the square bracket and thus we get,
\begin{eqnarray}
{\cal S} &\propto & 48 \sqrt{3} \frac{\eta_1}{\eta} \frac{1}{\eta^2} \frac{(k \eta)^{7/2}}{(k \eta)^{3/2}} \frac{ \left(1 - \sin(\frac{2k \eta}{\sqrt{3}})\right) }{\pi\, (k \eta)\,} \nonumber \\
 &\approx & \frac{1}{\eta^3} (k \eta) \propto \frac{1}{a^2}\, .
\end{eqnarray}
where we have used the fact that scale factor $\eta\propto a$ during RD epoch.
Therefore, for the modes ($k > k_{T1} $) which entered the horizon during SD era, source term will decay at most as $a^{-2}$ during the RD era. This implies that $\gamma_2(k)\leq 2 $ during RD era for these modes. 

Till now we have only considered the modes which entered the horizon during SD era and calculated the asymptotic values of $\gamma_1(k)$ and $\gamma_2(k)$. Now, for the modes which entered the horizon during RD era, scalar transfer function is same as obtained in the Ref.~\cite{Baumann:2007zm}. Hence, for such modes, the asymptotic values of $\gamma_2(k)$ will be the same as obtained in the Ref.~\cite{Baumann:2007zm}, which is $\gamma_2(k)\leq 4 $.

\end{document}